\documentclass[10pt,a4paper,useAMS,usenatbib]{article}
\usepackage{jcappub}
\usepackage[english]{babel}

\usepackage{booktabs}
\usepackage{fancyhdr}
\usepackage{amsfonts}
\usepackage{amsmath}
\usepackage{amssymb}
\usepackage{amsbsy}
\usepackage{txfonts}
\usepackage{multicol}
\usepackage{layout}
\usepackage{graphicx}
\usepackage{epstopdf}

\usepackage{times}
\usepackage{natbib}

\renewcommand{\d}{\mathrm{d}}

\newif\ifAMStwofonts
\AMStwofontstrue

\title{Constraining primordial non-Gaussianity with cosmological weak lensing: shear and flexion}

\author[a]{C. Fedeli}
\author[b]{M. Bartelmann}
\author[c,d,e]{L. Moscardini}

\affiliation[a]{Department of Astronomy, University of Florida, 211 Bryant Space Science Center, Gainesville, FL 32611-2055, USA}
\affiliation[b]{Zentrum f\"ur Astronomie, Universit\"at Heidelberg, Albert-\"Uberle-Stra\ss e 2, 69120 Heidelberg, Germany}
\affiliation[c]{Dipartimento di Astronomia, Universit\`a di Bologna, Via Ranzani 1, 40127 Bologna, Italy}
\affiliation[d]{INFN, Sezione di Bologna, Viale Berti Pichat 6/2, 40127 Bologna, Italy}
\affiliation[e]{INAF, Osservatorio Astronomico di Bologna, Via Ranzani 1, 40127 Bologna, Italy}

\emailAdd{cosimo.fedeli@astro.ufl.edu}

\abstract{We examine the cosmological constraining power of future large-scale weak lensing surveys on the model of the ESA planned mission \emph{Euclid}, with particular reference to primordial non-Gaussianity. Our analysis considers several different estimators of the projected matter power spectrum, based on both shear and flexion. We review the covariance and Fisher matrix for cosmic shear and evaluate those for cosmic flexion and for the cross-correlation between the two. The bounds provided by cosmic shear alone are looser than previously estimated, mainly due to the reduced sky coverage and background number density of sources for the latest \emph{Euclid} specifications. New constraints for the local bispectrum shape, marginalized over $\sigma_8$, are at the level of $\Delta f_\mathrm{NL} \sim 100$, with the precise value depending on the exact multipole range that is considered in the analysis. We consider three additional bispectrum shapes, for which the cosmic shear constraints range from $\Delta f_\mathrm{NL}\sim 340$ (equilateral shape) up to $\Delta f_\mathrm{NL}\sim 500$ (orthogonal shape). Also, constraints on the level of non-Gaussianity and on the amplitude of the matter power spectrum $\sigma_8$ are almost perfectly anti-correlated, except for the orthogonal bispectrum shape for which they are correlated. The competitiveness of cosmic flexion constraints against cosmic shear ones depends by and large on the galaxy intrinsic flexion noise, that is still virtually unconstrained. Adopting the very high value that has been occasionally used in the literature results in the flexion contribution being basically negligible with respect to the shear one, and for realistic configurations the former does not improve significantly the constraining power of the latter. Since the shear shot noise is white, while the flexion one decreases with decreasing scale, by considering high enough multipoles the two contributions have to become comparable. Extending the analysis up to $\ell_\mathrm{max} = 20,000$ cosmic flexion, while being still subdominant, improves the shear constraints by $\sim 10\%$ when added. However on such small scales the highly non-linear clustering of matter, the impact of baryonic physics, and the non-Gaussian part of the covariance matrix make any error estimation uncertain. By considering lower, and possibly more realistic, values of the flexion intrinsic shape noise results in flexion constraining power being a factor of $\sim 2$ better than that of shear, and the bounds on $\sigma_8$ and $f_\mathrm{NL}$ being improved by a factor of $\sim 3$ upon their combination.}
\keywords{gravitational lensing: weak $-$ large-scale structure of the Universe.}

\begin{document}
\maketitle

%%%%%%%%%%%%%%%%%%%%%%%%%%%%%%%%%%%%%%%%%%%%%%%%%%%
\section{Introduction}\label{sct:introduction}
%%%%%%%%%%%%%%%%%%%%%%%%%%%%%%%%%%%%%%%%%%%%%%%%%%%

In the forthcoming decade measurements of the fundamental parameters of cosmology will undergo a significant enhancement in precision, especially thanks to a number of space-based missions. The satellite \emph{Planck} \cite{AD11.1} has been taking data of the Cosmic Microwave Background (CMB henceforth) temperature and polarization fluctuations for about two years now, and their cosmological interpretation should soon be available. The X-ray observatory \emph{e}ROSITA \cite{PR07.1} is scheduled for launch in 2013, and will provide an all-sky catalog of galaxy clusters with unprecedent resolution and sensitivity for a survey of this size. Such a vast X-ray cluster sample is ideally suited for cosmology, as the uncertainties in the relations between cluster mass and observables can be treated as 'nuisance' parameters and marginalized over \cite{SA10.1}. In the more distant future, the \emph{Euclid}\footnote{http://www.euclid-ec.org} satellite \cite{LA11.1} is currently scheduled for launch in 2019, providing numerous opportunities for cosmological investigation with optical/NIR data: spectroscopy of a large number of galaxies over the almost entire extragalactic sky will allow measurements of the Baryon Acoustic Oscillation (BAO); spectroscopically, photometrically, and weak lensing-selected clusters of galaxies will provide a massive sample ideal for cosmological parameter estimation; the weak lensing map of almost half the sky will chart the matter power spectrum over unprecedent ranges of scales and redshifts, thus providing fundamental clues on the growth of matter perturbations. Planned by NASA, the Wide-Field InfraRed Survey Telescope (WFIRST, \cite{GR11.1}) is supposed to have mission objectives similar to those of \emph{Euclid}, although with more focus on supernova cosmology.

Given the enormous quantity of data that each one of these projects is expected to return, as well as the vast amount of resources that are invested, it is utterly important to explore different ways in which these products can be combined, and establish which data or data combinations are most promising for a certain cosmological application. This motivates us to review in this paper the constraining power of one of the most immediate science products for the \emph{Euclid} mission, that is the power spectrum of galaxy shapes. As galaxy shapes are affected by the light deflection operated by the Large-Scale Structure (LSS), their power spectrum is basically a projected (and weighted) version of the three-dimensional non-linear matter power spectrum, thus containing a wealth of cosmological information. The moment of the galaxy light distribution that is most commonly used is the ellipticity, since it provides a measure of the shear field \cite{BA01.1}. However, higher order lensing fields, affecting different light moments, exist and have been measured for individual galaxy clusters \cite{CA11.2,LE11.1}. Some of these fields, such as the flexion fields \cite{BA06.2}, will be measurable on cosmological scales by future weak lensing surveys such as \emph{Euclid} and WFIRST. In this work we assess the level at which cosmic shear and flexion, separately and in combination, can be used in order to constrain Primordial Non-Gaussianity (PNG), adopting \emph{Euclid} as our fiducial weak lensing survey. 

According to the standard picture for structure formation, the primordial density fluctuations that seeded today's LSS were produced during an early accelerated expansion phase of the Universe, dubbed inflation \cite{GU81.1}. The simplest inflationary model predicts almost Gaussian density fluctuations, while numerous more sophisticated models forecast arbitrary deviations from Gaussianity. If occurring, PNG would have a significant effect both on CMB and cosmic structures. As a matter of fact, the issue of constraining deviations from primordial Gaussianity by using structure formation and evolution has recently attracted much attention in the literature, with efforts directed towards the abundance of dark matter halos \cite{MA00.2,VE00.1,MA04.1,LO08.1,GR09.1,MA10.1}, halo biasing \cite{DA08.1,MA08.1,MC08.1,FE09.1,FE11.1,NA11.1}), galaxy bispectrum \cite{SE07.2,JE09.1}, mass density distribution \cite{GR08.2} and topology \cite{MA03.2,HI08.2}, integrated Sachs-Wolfe effect \cite{AF08.1,CA08.1,TA12.1}, Ly$\alpha$ flux from low-density intergalactic medium \cite{VI09.1}, $21-$cm fluctuations \cite{CO06.2,PI07.1}, reionization \cite{CR09.1}, and weak lensing peak counts \cite{MA11.3,MA11.4}. PNG is one of the very few handle that we have on the pre-recombination universe, and it is hence important to study it in detail and with different probes.

Cosmic shear from future weak lensing surveys has been investigated as a tool for probing PNG in \cite{FE10.1} (see also \cite{GI11.1}). Here we review those constraints with a more complete statistical analysis and by using the latest \emph{Euclid} specifications for the galaxy number density, the intrinsic ellipticity noise, the sky coverage, and the source redshift distribution. In addition, we consider cosmic flexion as an alternative channel of cosmological information. The intrinsic shape noise for flexion has not been reliably measured yet. Values adopted in the literature are usually very high, hence they are expected to result in flexion giving only a subdominant contribution to the overall \emph{Euclid} constraining power. However, given the large uncertainty on the flexion noise and the fact this noise is scale-dependent, it is important to investigate if the flexion contribution is indeed negligible and if there are configurations for which it can become relevant. We dedicate the Appendix \ref{sct:appendix} to the calculations of covariances and Fisher matrices for cosmic shear, cosmic flexion, and their cross-correlation, the latter two of which, to the best of our knowledge, have not been reported elsewhere. The main body of this paper is organized as follows. In Section \ref{sct:ng} we review how PNG is parametrized and in Section \ref{sct:impact} we summarize the effect thereof on halo mass function, bias, and internal structure. In Section \ref{sct:halomodel} we describe the halo model, a physically motivated framework that we adopted in order to estimate the fully non-linear matter and lensing power spectra in non-Gaussian cosmologies. In Section \ref{sct:constraints} we show our results concerning joint constraints on the level of PNG and the amplitude of the matter power spectrum from cosmic shear and cosmic flexion. Section \ref{sct:conclusions} is dedicated to a discussion of the results and a summary of the conclusions.

Throughout this work we adopted as a fiducial cosmological model the one suggested by the latest analysis of the WMAP data \cite{KO11.1}, bearing $\Omega_{\mathrm{m},0}=0.272$, $\Omega_{\Lambda,0} = 1-\Omega_{\mathrm{m},0}$, $\Omega_{\mathrm{b},0}=0.046$, $H_0 = h100$ km s$^{-1}$ Mpc$^{-1}$ with $h=0.704$, and a matter power spectrum normalized by setting $\sigma_8=0.809$.

%%%%%%%%%%%%%%%%%%%%%%%%%%%%%%%%%%%%%%%%%%%%%%%%%%%
\section{Non-Gaussian cosmologies}\label{sct:ng}
%%%%%%%%%%%%%%%%%%%%%%%%%%%%%%%%%%%%%%%%%%%%%%%%%%%

%%%%%%%%%%%%%%%%%%%%%%%%%%%%%%%%%%%%%%%%%%%%%%%%%%%
\subsection{General}
%%%%%%%%%%%%%%%%%%%%%%%%%%%%%%%%%%%%%%%%%%%%%%%%%%%

As mentioned in the Introduction, extensions of the most standard model of inflation \cite{ST79.1,GU81.1,LI82.1} can produce substantial deviations from a Gaussian distribution of primordial density and potential fluctuations (see \cite{BA04.1,CH10.1} for recent reviews). The amount and shape of these deviations depend critically on the kind of non-standard inflationary model that one has in mind, as will be detailed later on.

A particularly convenient (although not unique) way to describe generic deviations from Gaussian initial conditions consists in writing the gauge-invariant Bardeeen's potential $\Phi$ as the sum of a Gaussian random field and a quadratic correction \cite{SA90.1,GA94.1,VE00.1,KO01.1}, according to

\begin{equation}\label{eqn:ng}
\Phi = \Phi_\mathrm{G} + f_\mathrm{NL} * \left( \Phi_\mathrm{G}^2 - \langle \Phi_\mathrm{G}^2 \rangle \right)~.
\end{equation}
The parameter $f_\mathrm{NL}$ in Eq. (\ref{eqn:ng}) determines the amplitude of non-Gaussian deviations, and it is in general dependent on the scale. The symbol $*$ denotes convolution between functions, and reduces to standard multiplication upon constancy of $f_\mathrm{NL}$. In the following we adopted the LSS convention (as opposed to the CMB convention, see \cite{AF08.1,CA08.1,PI09.1,GR09.1}) for the definition of the fundamental parameter $f_\mathrm{NL}$. According to this, the primordial value of $\Phi$ has to be linearly extrapolated at $z = 0$, and as a consequence the constraints given on $f_\mathrm{NL}$ by the CMB have to be raised by $\sim 30$ per cent to comply with this paper's convention (see also \cite{FE09.1} for a concise explanation).

If $f_\mathrm{NL} \ne 0$ the potential $\Phi$ is a random field with a non-Gaussian probability distribution. Therefore, the field itself cannot be described by the linear power spectrum $P_{\Phi,\mathrm{L}}(\boldsymbol k) = Ak^{n-4}$ alone, rather higher order statistics are needed. In many circumstances the dominant higher order statistic is the bispectrum $B_\Phi(\boldsymbol k_1,\boldsymbol k_2,\boldsymbol k_3)$. The bispectrum is the Fourier transform of the three-point correlation function $\langle \Phi(\boldsymbol k_1)\Phi(\boldsymbol k_2)\Phi( \boldsymbol k_3) \rangle$ and it can hence be implicitly defined as

\begin{equation}
\langle \Phi(\boldsymbol k_1)\Phi(\boldsymbol k_2)\Phi(\boldsymbol k_3) \rangle \equiv (2\pi)^3\delta_\mathrm{D}\left( \boldsymbol k_1+\boldsymbol k_2+\boldsymbol k_3 \right) B_\Phi(\boldsymbol k_1, \boldsymbol k_2, \boldsymbol k_3)~,
\end{equation}
where $\delta_\mathrm{D}$ is the Dirac delta distribution and the angular brackets indicate an ensemble average.

Since different inflationary models predict different non-Gaussian shapes, that is different ways in which the potential bispectrum depends on its three arguments, understanding the PNG shape is of fundamental importance in order to pinpoint the physics of the early Universe. In the standard inflationary scenario the early accelerated expansion is accounted for by a slowly-rolling scalar field, the inflaton. In the present work we considered four different shapes of the potential bispectrum, arising from different modifications of this standard scenario. They are all generically described in the following. For further details we refer the reader to \cite{FE11.1}.

%%%%%%%%%%%%%%%%%%%%%%%%%%%%%%%%%%%%%%%%%%%%%%%%%%%
\subsection{Shapes}
%%%%%%%%%%%%%%%%%%%%%%%%%%%%%%%%%%%%%%%%%%%%%%%%%%%

\subsubsection{Local shape}

The standard single-field inflationary scenario generates negligibly small deviations from Gaussianity. These deviations are commonly said to be of the local shape, and the related bispectrum of the Bardeen's potential is maximized for \emph{squeezed} configurations, where one of the three wavevectors has a much smaller magnitude than the other two. In this case the parameter $f_\mathrm{NL}$ must be a constant, and it is expected to be of the same order of the slow-roll parameters \cite{FA93.1}, that are in turn of order unity. 

However non-Gaussianities of the local shape can also be generated in the case in which an additional light scalar field, different from the inflaton, contributes to the observed curvature perturbations \cite{BA04.2}. This happens, for instance, in curvaton models \cite{SA06.1,AS07.1} or in multi-field models \cite{BA02.2,BE02.1}. In this case the parameter $f_\mathrm{NL}$ is allowed to be substantially different from zero. The best constraints on the level of non-Gaussianity for a local bispectrum shape come from the WMAP$-7$ data \cite{KO11.1}, and at $95\%$ Confidence Level (CL) amount to $-10 < f_\mathrm{NL}<74$.

\subsubsection{Equilateral shape}

In some inflationary models the kinetic term of the inflaton Lagrangian is not standard, containing higher-order derivatives of the field itself. One significant example of this is the DBI model (\cite{AL04.1,SI04.1}, see also \cite{AR04.1,SE05.1,LI08.1}). In this case the primordial bispectrum is maximized for configurations where the three wavevectors have approximately the same amplitude. A template for this equilateral bispectrum can be found in \cite{CR07.1}.

Unlike the local shape, for the equilateral case there is no theoretical prescription against a running of the $f_\mathrm{NL}$ parameter with the scale, and in fact many authors introduce such a running in order to enhance deviations from Gaussianity at scales smaller than those probed by CMB studies. In this case however we opted for not introducing such a running, in order to directly compare pure bispectrum shapes. Constraints on $f_\mathrm{NL}$ in this case have also been given by the WMAP team \cite{KO11.1}, and correspond to $-214 < f_\mathrm{NL}<266$ at $95\%$ CL.

\subsubsection{Enfolded shape}

For deviations from Gaussianity evaluated in the regular Bunch-Davies vacuum state, the primordial potential bispectrum is of local or equilateral shape, depending on whether or not higher-order derivatives play a significant role in the evolution of the inflaton field. If the Bunch-Davies vacuum hypothesis is dropped, the resulting bispectrum is maximal for \emph{squashed} configurations \cite{CH07.1,HO08.1}. Authors in \cite{ME09.1} found a template that describes very well the properties of this enfolded-shape bispectrum.

Constraints on the level of PNG for the enfolded-shaped bispectrum have not been given in \cite{KO11.1}. However, relatively tight bounds can be estimated from the correlation functions of LSS tracers. Specifically, authors in \cite{XI11.1} give $-12 < f_\mathrm{NL}<358$ (CMB convention) at the $2\sigma$ CL.

\subsubsection{Orthogonal shape}

A shape of the bispectrum can be constructed that is nearly \emph{orthogonal} (with respect to a suitably defined scalar product) to both the local and equilateral forms \cite{SE10.1}. This kind of bispectrum is peaked both on equilateral and squashed configurations, with opposite signs. Constraints on the level of non-Gaussianity compatible with the CMB in the orthogonal scenario were also given by the WMAP team \cite{KO11.1}, corresponding to $-410 < f_\mathrm{NL}<6$ at $95\%$ CL.

Similarly to the equilateral case, for enfolded and orthogonal shapes it would be allowed to introduce a running of the parameter $f_\mathrm{NL}$ with scale, which we decided not to include. However it should be stressed that, differently from the equilateral shape, in the two latter cases there is no first principle that can guide one in the choice of a particular kind of running, and until now no work has addressed the problem of a running for these shapes \cite{FE09.2,FE10.2}.

%%%%%%%%%%%%%%%%%%%%%%%%%%%%%%%%%%%%%%%%%%%%%%%%%%%
\section{Impact of PNG on structure formation}\label{sct:impact}
%%%%%%%%%%%%%%%%%%%%%%%%%%%%%%%%%%%%%%%%%%%%%%%%%%%

PNG produces modifications in the statistics of density peaks, resulting in differences in the mass function, bias, and internal structure of dark matter halos. Since all these factors enter in the modeling of the LSS, we summarize in the following how these modifications can be parametrized.

%%%%%%%%%%%%%%%%%%%%%%%%%%%%%%%%%%%%%%%%%%%%%%%%%%%
\subsection{Mass function}
%%%%%%%%%%%%%%%%%%%%%%%%%%%%%%%%%%%%%%%%%%%%%%%%%%%

We modeled the non-Gaussian contribution to the number counts of dark matter halos according to the prescription presented in \cite{LO08.1}. These authors gave a useful expression for the Press \& Schechter \cite{PR74.1} mass function following from non-Gaussian initial conditions, $n_\mathrm{PS}(M,z)$. Since the corresponding mass function computed with Gaussian initial conditions, $n^\mathrm{(G)}_\mathrm{PS}(M,z)$, is well known, we can define a correction factor $\mathcal{R}(M,z)\equiv n_\mathrm{PS}(M,z)/n^\mathrm{(G)}_\mathrm{PS}(M,z)$. Under the assumption that  the effect of PNG on the mass function is independent of the prescription adopted to describe the mass function itself, it follows that the non-Gaussian mass function computed according to an arbitrary prescription, $n(M,z)$, can be related to its Gaussian counterpart through

\begin{equation}
n(M,z) = \mathcal{R}(M,z)~n^\mathrm{(G)}(M,z)~.
\end{equation}

In order to evaluate $n_\mathrm{PS}(M,z)$, and hence $\mathcal{R}(M,z)$, authors in \cite{LO08.1} performed an Edgeworth expansion \cite{BL98.1} of the probability distribution for the smoothed density fluctuations field, truncating it at the linear term in $\sigma_M$. One function that turns out to be needed in the subsequent formula is the reduced skewness $S_3(M)\equiv f_\mathrm{NL}~\mu_3(M)/\sigma_M^4$ of the non-Gaussian distribution, where the skewness $\mu_3(M)$ can be computed as

\begin{equation}
\mu_3(M) = \int_{\mathbb{R}^9}  \frac{\mathrm{d}^3 \boldsymbol k_1\mathrm{d}^3 \boldsymbol k_2 \mathrm{d}^3 \boldsymbol k_3}{(2\pi)^9}\mathcal{M}_R(k_1) \mathcal{M}_R(k_2) \mathcal{M}_R(k_3) \langle\Phi(\boldsymbol k_1)\Phi(\boldsymbol k_2)\Phi( \boldsymbol k_3)\rangle~.
\end{equation}
The function $\mathcal{M}_R(k)$ relates the density fluctuations smoothed on some scale $R$ (corresponding to the mass $M$) to the respective peculiar potential,

\begin{equation}
\mathcal{M}_R(k) \equiv \frac{2}{3}\frac{T(k)k^2c^2}{H_0^2\Omega_{\mathrm{m},0}}W_R(k)~,
\end{equation}
where $T(k)$ is the matter transfer function and $W_R(k)$ is the Fourier transform of the top-hat window function. In this work we adopted the Bardeen et al. \cite{BA86.1} matter transfer function, with the shape factor correction presented in \cite{SU95.1}. This reproduces fairly well the more sophisticated recipe of Eisenstein \& Hu \cite{EI98.1} except for the presence of the BAO, that anyway is not of interest here. 

For the reference Gaussian mass function we adopted the Sheth \& Tormen prescription \cite{SH02.1} (see \cite{JE01.1,WA06.1,TI08.1} for alternative prescriptions). It is worth noting that other approaches for computing the non-Gaussian correction to the halo number counts exist (see for instance \cite{MA00.2}). These alternative approaches give results that are in broad agreement with those of \cite{LO08.1}, and have been collectively shown to provide a good representation of the mass function measured in cosmological simulations \cite{GR09.1,WA10.1}.

%%%%%%%%%%%%%%%%%%%%%%%%%%%%%%%%%%%%%%%%%%%%%%%%%%%
\subsection{Halo bias}\label{sct:ngbias}
%%%%%%%%%%%%%%%%%%%%%%%%%%%%%%%%%%%%%%%%%%%%%%%%%%%

In this Subsection we describe how the bias of dark matter halos gets modified by PNG. The fact that deviations from primordial Gaussianity induce a scale-dependence on the large-scale halo bias has been first noticed by \cite{DA08.1}. Later, authors in \cite{MA08.1} used the peak-background split formalism in order to provide a semi-analytic expression for the non-Gaussian bias, hence allowing to place and forecast stringent constraints on PNG by using the correlation functions of LSS tracers \cite{CA08.1,AF08.1,SL08.1,VE09.1,XI11.1}.

The non-Gaussian halo bias can be written in a relatively straightforward way in terms of its Gaussian counterpart as \cite{CA10.1}

\begin{equation}
b(M,z,k) = b^\mathrm{(G)}(M,z) + \beta_R(k)\sigma_M^2\left[ b^\mathrm{(G)}(M,z)-1 \right]^2,
\end{equation}
where the function $\beta_R(k)$ encapsulates all the scale dependence of the non-Gaussian correction to the bias. It can be written as

\begin{equation}
\beta_R(k) = \frac{1}{8\pi^2\sigma_M^2\mathcal{M}_R(k)} \int_0^{+\infty} \mathrm{d}\zeta~\zeta^2\mathcal{M}_R(\zeta)
\left[ \int_{-1}^1\mathrm{d}\mu \mathcal{M}_R\left(\alpha\right) \frac{B_\Phi\left( \zeta,\alpha,k \right)}{P_\Phi(k)} \right]~,
\end{equation}
with $\alpha^2 \equiv k^2 + \zeta^2 + 2 k\zeta\mu$. In the simple case of local bispectrum shape it can be shown that the function $\beta_R(k)$ should scale as $\propto k^{-2}$ at large scales, so that a substantial boost (for positive $f_\mathrm{NL}$) in the halo bias is expected at those scales. For the Gaussian halo bias we adopted the Sheth, Mo \& Tormen \cite{SH01.1} prescription.

We emphasize that the semi-analytic prescription by \cite{MA08.1} has also been extensively tested against numerical simulations, giving an overall good agreement \cite{GR09.1,WA12.1}.

%%%%%%%%%%%%%%%%%%%%%%%%%%%%%%%%%%%%%%%%%%%%%%%%%%%
\subsection{Halo internal structure}\label{sct:structure}
%%%%%%%%%%%%%%%%%%%%%%%%%%%%%%%%%%%%%%%%%%%%%%%%%%%

Since PNG affects the timing of structure formation, it is expected to have some sort of impact on the internal structure of dark matter halos. However, to date there are only a handful of works exploring this matter. After the pioneering study by \cite{AV03.1}, more recently this issue has been addressed in passing by \cite{SM11.1} through numerical simulations, and by \cite{DA11.1} via semi-analytic considerations. These authors assume the average dark matter halo to be well represented by a Navarro, Frenk \& White \cite{NA96.1} (NFW henceforth) density profile also in non-Gaussian cosmologies. However the concentration of the profile turns out to be enhanced in models with positive skewness and depressed in models with negative skewness. The results of the semi-analytic model of \cite{DA11.1} fairly agree with the simulations by \cite{SM11.1}, showing that the halo concentration is increased (decreased) by $\sim 4-10\%$ for a local bispectrum shape with $f_\mathrm{NL} = +100$ ($f_\mathrm{NL} = -100$, CMB convention). The effect is mildly dependent on mass and redshift, with more massive objects placed at higher redshift being the most affected by PNG. The semi-analytic model by \cite{DA11.1} is however quite cumbersome. A simplified model has been proposed by \cite{OG09.2}. The latter reproduces well the results presented in \cite{DA11.1} at low redshift and on galaxy group-scales, but it somewhat underestimates them for large masses/high redshifts. 

In this work we assumed average dark matter halos to be well represented by NFW density profiles, however we adopted the same concentration prescription for all cosmological models, which matches the results of $\Lambda$CDM cosmological simulations (see \cite{FE10.1} for more details). The reason for this choice is twofold: $i)$ the previous studies are limited to the local bispectrum shape, while in this work we explored a variety of non-Gaussian shapes. In principle the model by \cite{DA11.1} is of general application, however its agreement with cosmological simulations has not been tested so generally; $ii)$ in general we limited our analysis to relatively large scales, where the impact of the very inner structure of dark matter halos can be considered irrelevant. In the few cases in which it is not (that will be clearly indicated), one should keep in mind that the impact of baryonic physics and the non-Gaussian parts of the covariances (see Appendix \ref{sct:appendix} for details) constitute more of an uncertainty in the statistical analysis. We return on this point further below.

%%%%%%%%%%%%%%%%%%%%%%%%%%%%%%%%%%%%%%%%%%%%%%%%%%%
\section{Modeling the large-scale structure}\label{sct:halomodel}
%%%%%%%%%%%%%%%%%%%%%%%%%%%%%%%%%%%%%%%%%%%%%%%%%%%

%%%%%%%%%%%%%%%%%%%%%%%%%%%%%%%%%%%%%%%%%%%%%%%%%%%
\subsection{Halo model}
%%%%%%%%%%%%%%%%%%%%%%%%%%%%%%%%%%%%%%%%%%%%%%%%%%%

We represented the non-linear matter power spectrum by making use of the halo model \cite{MA00.3,SE00.1}. This is a physical framework that allows the description of the correlation of dark matter particles as well as that of different tracers of the LSS such as galaxies ad galaxy clusters. It is based on the fact that the power spectrum of \emph{particles} (either dark matter particles or tracers) is given by the sum of two contributions: particle pairs residing in the same structure, and particle pairs residing in two different structures. Accordingly, the dark matter power spectrum can be written as the sum of two contributions,

\begin{equation}
P_\mathrm{m}(k,z) = P_{\mathrm{m},1}(k,z)+P_{\mathrm{m},2}(k,z).
\end{equation}
The first term is named $1-$halo term, and dominates on very small scales, while the second is the $2-$halo term, and it is dominant at large scales. It is easy to understand that the $2-$halo term should incorporate the bias of dark matter halos, since it represents the clustering of particle pairs residing in separated structures \cite{CO02.2}.

The two contributions to the power spectrum of dark matter particles can be written as

\begin{equation}
P_{\mathrm{m},1}(k,z) = \int_0^{+\infty} \mathrm{d}M~n(M,z) \left[\frac{\hat{\rho}(M,z,k)}{\rho_\mathrm{m}}\right]^2
\end{equation}
and

\begin{equation}
P_\mathrm{m,2}(k,z) = \left[\int_0^{+\infty} \mathrm{d}M~n(M,z) b(M,z,k) \frac{\hat{\rho}(M,z,k)}{\rho_\mathrm{m}}\right]^2P_\mathrm{m,L}(k,z)~.
\end{equation}
In the previous pair of equations, $\rho_\mathrm{m}$ represents the comoving matter density, $P_\mathrm{m,L}(k,z)$ is the linear matter power spectrum, $n(M,z)$ is the mass function, and $b(M,z,k)$ is the halo bias, where we inserted an additional scale dependence in order to account for the effect of primordial non-Gaussianity (see Section \ref{sct:ngbias}). The function $\hat{\rho}(M,z,k)$ represents the Fourier transform of the average density profile of dark matter halos,

\begin{equation}
\hat{\rho}(M,z,k) = \int_0^{R_\mathrm{v}}r^2 \mathrm{d}r~\rho(M,z,r) \frac{\sin(kr)}{kr}~.
\end{equation}
Since for $\rho(M,z,k)$ we adopted a NFW shape, its Fourier transform can be computed analytically as described in \cite{SC01.1,RU08.2}. Because the integral in the previous equation is evaluated up to the virial radius $R_\mathrm{v}$ of the structure, it is easy to see that in the large scale limit $\hat{\rho}(M,z,k)$ converges to the virial mass of the halo.

There are two assumptions underlying the halo model representation of the dark matter power spectrum. The first one is that all the matter in the Universe is contained within halos of some mass, that is

\begin{equation}
\int_0^{+\infty} M \mathrm{d}M~n(M,z) = \rho_\mathrm{m}~.
\end{equation}
As can be verified, this assumption is fulfilled by the Press \& Schechter \cite{PR74.1} and Sheth \& Tormen \cite{SH02.1} mass functions, of which we are considering the latter. The second assumption is that the fully non-linear power spectrum should converge to the linear one at very large scales. In other words, this implies the non-trivial constraint

\begin{equation}
\int_0^{+\infty} M \mathrm{d}M~n(M,z) b(M,z,k) = \rho_\mathrm{m}
\end{equation}
for very large scales. This constraint has to be explicitly enforced, since the standard halo bias formulations do not satisfy it automatically. For this and other details on the practical implementation of the halo model we refer the interested reader to \cite{FE10.1}.

As a last note, in \cite{SM11.1} the authors pointed out that a better agreement of the non-linear power spectrum estimated by the halo model with numerical simulations may be obtained by considering the so-called halo exclusion term. This consists in an integral to be subtracted by the $2-$halo contribution, taking into account that dark matter halos are considered as hard spheres that cannot overlap with one another. We ignored this possible improvement, as what matters in our statistical analysis is only the relative change induced by PNG on the non-linear matter power spectrum.

%%%%%%%%%%%%%%%%%%%%%%%%%%%%%%%%%%%%%%%%%%%%%%%%%%%
\subsection{Cosmological weak lensing}
%%%%%%%%%%%%%%%%%%%%%%%%%%%%%%%%%%%%%%%%%%%%%%%%%%%

An effective way to probe the large-scale matter distribution in the Universe is offered by the gravitational deflection of light. Specifically, the weak distortion of images of background galaxies as their light passes through the LSS allows to map the total matter distribution up to the source redshift, projected along the line of sight. Therefore, the power spectrum of lensing-induced distortions can be thought as a projection of the matter power spectrum described in the previous Subsection, weighted with a function that depends on the redshift distribution of background sources.

The quantity relevant for describing the correlation function of lensing-induced distortions is the convergence power spectrum, that can be related to the fully non-linear matter power spectrum through \cite{KA92.1,BA01.1}

\begin{equation}\label{eqn:convergenceps}
P_\kappa(\ell) = \frac{9H_0^4\Omega_{\mathrm{m},0}^2}{4c^4} \int_0^{\chi_\mathrm{h}} \mathrm{d}\chi \frac{W^2(\chi)}{a^2(\chi)} P_\mathrm{m}\left[\frac{\ell}{f_K(\chi)},\chi \right]~,
\end{equation}
where $\chi=\chi(z)$ is the comoving distance out to redshift $z$, the scale factor $a(\chi)$ is normalized such that $a(0)=1$, and $f_K(\chi)$ is the comoving angular diameter distance, which in general depends on the curvature of the Universe. In this work we considered flat universes only, so that $f_K(\chi) = \chi$, however we keep the formalism as general as possible. As mentioned, the weight function $W(\chi)$ is related to source redshift distribution $g(z)$ by

\begin{equation}
W(\chi) = \int_\chi^{\chi_\mathrm{h}} \mathrm{d}\chi' g(\chi')\frac{f_K(\chi-\chi')}{f_K(\chi')}~.
\end{equation}
Both the integrals above formally extend up to the comoving horizon distance $\chi_\mathrm{h}$, however the source redshift distribution declines to zero well before that, so that the integral can be effectively thought as extending out to infinity.

\begin{figure}
\centering
\includegraphics[width=0.49\hsize]{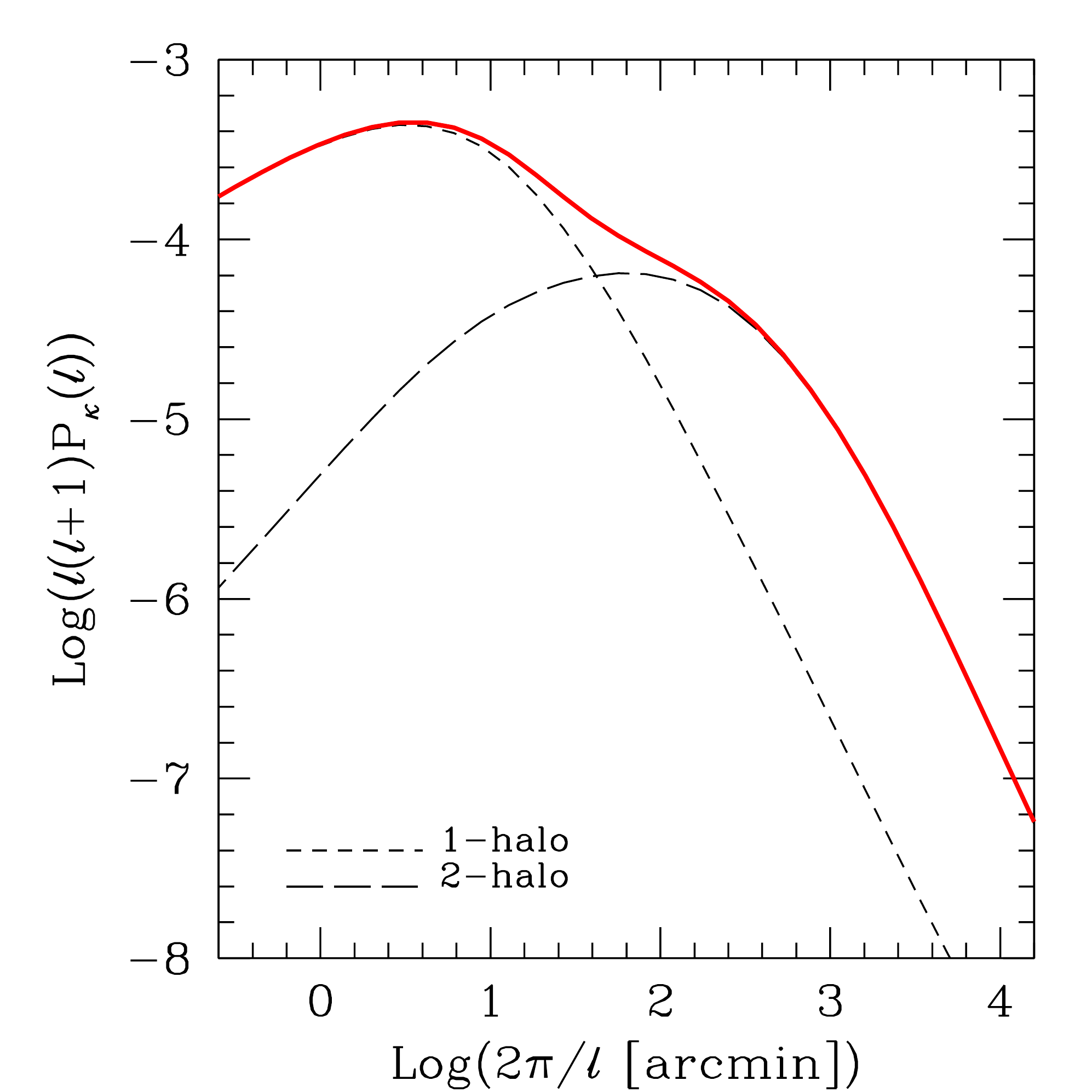}
\includegraphics[width=0.49\hsize]{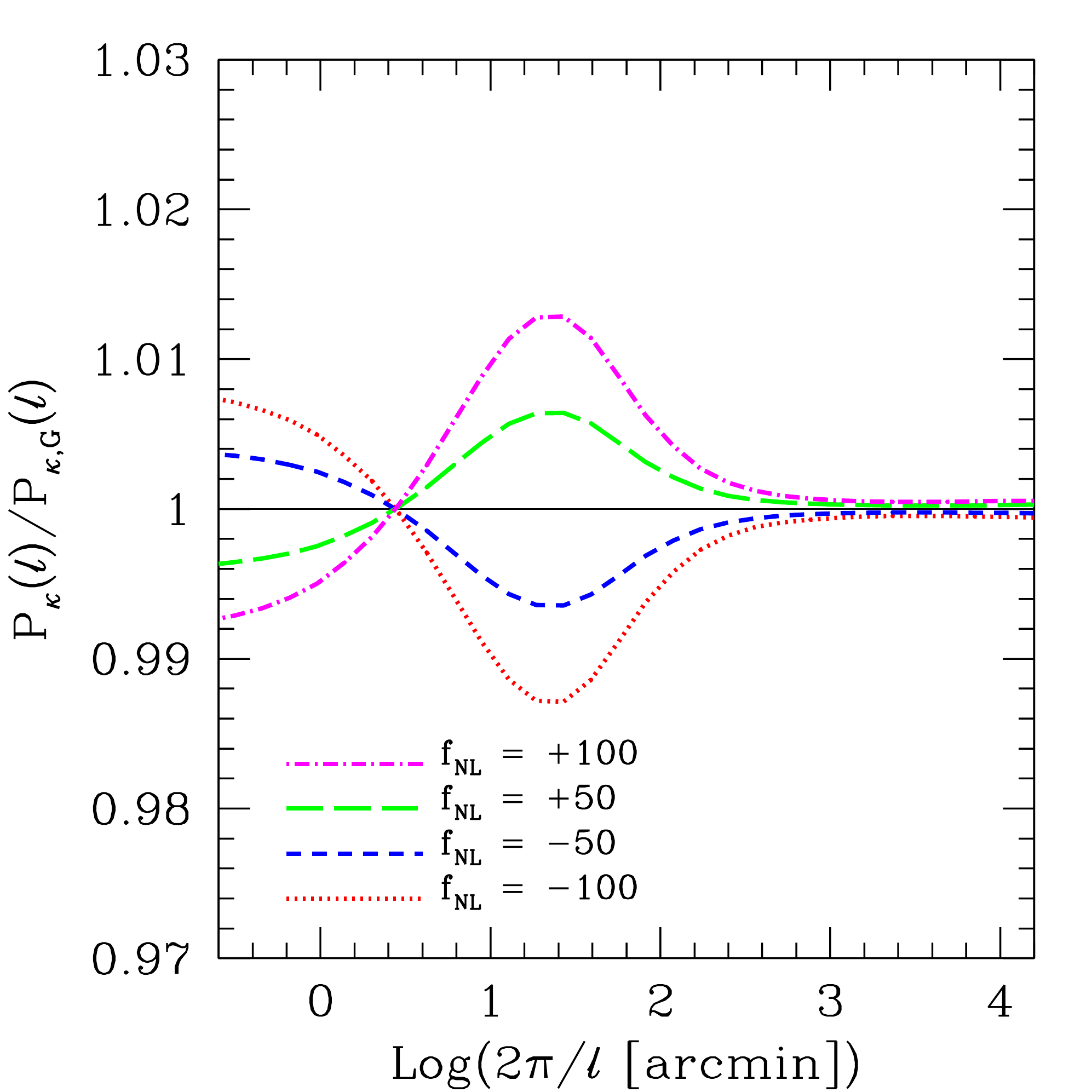}
\caption{\emph{Left panel}. The convergence power spectrum as a function of the angular scale for the fiducial Gaussian cosmological model. The solid red line represents the total spectrum, while the two black dashed lines refer to the $1-$halo and $2-$halo contributions. \emph{Right panel}. The convergence power spectra computed for the local shape non-Gaussian bispectrum with different values of the parameter $f_\mathrm{NL}$, as labeled. All spectra are divided by the convergence power spectrum in the reference Gaussian cosmology.}
\label{fig:weakLensing}
\end{figure}

It should be kept in mind that the previous equation for the convergence power spectrum is not generally applicable, rather it makes use of the Limber approximation for flat-sky quantities. This approximation is valid only as long as we consider sufficiently small patches of the sky. Specifically, it has been shown by \cite{JE09.1} that the accuracy of the Limber approximation for the convergence power spectrum is at the level of $1\%$ for $\ell > 10$, corresponding to angular scales smaller than $2\times 10^{3}$ arcmin.

The convergence power spectrum cannot be measured directly, but it can be estimated by measuring the correlation functions of other (observable) lensing fields. The field that has been mainly used in the past for this purpose is the shear $\gamma$, because it affects only the ellipticity of background sources and it is hence relatively easy to measure. The two flexion fields, which affect higher-order moments of the galaxy light distribution have been measured recently in galaxy cluster fields, and will certainly be possible to measure them on cosmological scales with \emph{Euclid}. In this work we ignored the second flexion because it is trickier to measure, and focused instead only on the first flexion $\varphi$, which induces an arc-like distortion on background sources \cite{SC08.2}. In the Appendix \ref{sct:appendix} we summarize how unbiased estimators of the convergence power spectrum can be constructed which are based on both shear (cosmic shear) and flexion (cosmic flexion) measurements, as well as their cross-correlation. We also show computations of the covariances and Fisher matrices for these estimators, which will be needed in the subsequent statistical analysis.

In order to refer to a specific case, in this work we adopted the guidelines set up for the planned \emph{Euclid} weak lensing survey, as they are stated in the \emph{Euclid Red Book} \cite{LA11.1}. Different specifications are provided for different survey configurations. Here we chose the configuration covering a solid angle of $\Omega = 15,000$ deg$^2$ over the extragalactic sky ($f_\mathrm{sky} = 0.364$). In this case the source redshift distribution reads

\begin{equation}
g(z) = \frac{\beta}{z_\star\Gamma\left[(\alpha+1)/\beta\right]} \left(\frac{z}{z_\star}\right)^\alpha \exp\left[ -\left( \frac{z}{z_\star} \right)^\beta \right]~,
\end{equation}
with $z_\star = 0.6374$, $\alpha = 2$, and $\beta = 1.5$. The corresponding median source redshift is $z_\mathrm{m} = 0.9$. 

As can be seen from the Fisher matrices reported in Eqs. (\ref{eqn:fisher_shear}), (\ref{eqn:fisher_flexion}), and (\ref{eqn:fisher_cross}), a number of other parameters need to be specified in order to perform a statistical analysis. The first one is the average number density of background sources for which a shape measurement is feasible. The \emph{Euclid} minimum requirement is $\bar n = 30$ arcmin$^{-2}$, while its goal is $\bar n = 40$ arcmin$^{-2}$. We adopted the former in this work. Also, we have to select the width of the multipole band where averages of the convergence power spectrum estimators are taken (see Eq. (\ref{eqn:cov_shear4}) and the discussion that follows). Consistently with the majority of previous works, we used $\Delta\ell = 1$ hereafter. The most delicate parameters to be selected are probably the intrinsic shape noises for the shear of background sources $\sigma_\gamma$, for the flexion $\sigma_\varphi$, and for their cross-correlation $\sigma_x$ (see Appendix \ref{sct:appendix} for formal definitions), since they enter squared in the respective Fisher matrices. The value of the intrinsic shear noise suggested for \emph{Euclid} is $\sigma_\gamma = 0.3$ (see however \cite{ZH09.1} for a different choice), but there are no standard values for $\sigma_\varphi$ and $\sigma_x$. As a reference value we adopted the one suggested by \cite{CA11.1}, that is $\sigma_\varphi = 6.19\times 10^{3}$. This choice is discussed in more detail further below, where we also show the consequences of adopting a substantially smaller value. Lacking any evidence to the contrary, here and in the following we considered $\sigma_x = (\sigma_\gamma\sigma_\varphi)^{1/2}$. This implies $\sigma_x = 43.1$ for our current choice of parameters. With these assumptions, and considering the different scale dependencies of the flexion shot noise and the shear shot noise (compare Eq. \ref{eqn:covariance_shear} with Eq. \ref{eqn:covariance_flexion}), the former becomes smaller than the latter for $\ell \gtrsim 20,000$ \cite{PI10.1}.

%%%%%%%%%%%%%%%%%%%%%%%%%%%%%%%%%%%%%%%%%%%%%%%%%%%
\section{Constraints on PNG}\label{sct:constraints}
%%%%%%%%%%%%%%%%%%%%%%%%%%%%%%%%%%%%%%%%%%%%%%%%%%%

%%%%%%%%%%%%%%%%%%%%%%%%%%%%%%%%%%%%%%%%%%%%%%%%%%%
\subsection{Modifications to the convergence power spectrum}
%%%%%%%%%%%%%%%%%%%%%%%%%%%%%%%%%%%%%%%%%%%%%%%%%%%

In this Subsection we summarize the impact of PNG on the convergence power spectrum. Further details on this topic can be found in \cite{FE10.1}. For reference, in the left panel of Figure \ref{fig:weakLensing} we show the convergence power spectrum evaluated in the reference Gaussian cosmology as a function of angular scale. We report the contributions of the $1-$halo and the $2-$halo terms of the non-linear matter power spectrum. As expected, the former dominates on very small angular scales, while the latter is more important at very large scales. Our subsequent analysis shall be limited at multipoles above $\ell_\mathrm{min} = 50$ ($\theta \lesssim 430$ arcmin). While the \emph{Euclid Red Book} recommends the value $\ell_\mathrm{min} = 5$, we preferred to adopt the former value in order for the Limber approximation used above to be valid with high accuracy.

In the right panel of Figure \ref{fig:weakLensing} we report the impact of PNG on the convergence power spectrum. For illustrative purposes we only consider non-Gaussianity with local shape, and a range of $f_\mathrm{NL}$ values that brackets the current best constraints. As can be seen, the convergence power spectrum shows an increase at intermediate angular scales for positive skewness, and a decrease for negative ones. The peak of the difference is reached at $\sim 30$ arcmin angular scale, and it is somewhat larger than $\sim 1\%$ for $f_\mathrm{NL} = \pm 100$. 
Since the shear shot noise is white, while those of flexion and of their cross-correlation decrease toward small scales (see the covariances computed in Eqs. \ref{eqn:covariance_shear}, \ref{eqn:covariance_flexion}, and \ref{eqn:covariance_cross}), it is expected that flexion constraints on PNG would improve much faster than shear constraints when including smaller and smaller scales. In the next Subsection we show how this expectation is verified.

Note that the behavior depicted in the right panel of Figure \ref{fig:weakLensing} is, for small angular scales, different from the results presented in, e.g., \cite{HI12.1}. That work is based on numerical $n-$body simulations, and the modifications to the cosmic shear correlation function do not show any crossover. This is ascribable to the effect of PNG on the inner structure of dark matter halos, which we ignored, and makes our subsequent analysis conservative modulo the effect of baryons.

%%%%%%%%%%%%%%%%%%%%%%%%%%%%%%%%%%%%%%%%%%%%%%%%%%%
\subsection{Fisher matrix analysis}
%%%%%%%%%%%%%%%%%%%%%%%%%%%%%%%%%%%%%%%%%%%%%%%%%%%

We performed a Fisher matrix analysis in order to forecast the constraints that the convergence power spectrum, estimated either via shear, flexion, or their cross-correlation, can put on the level of PNG, depending on the shape of the primordial bispectrum. As already mentioned, the detailed calculations for these Fisher matrices are collected in Appendix \ref{sct:appendix}, to which we refer the interested reader. The fiducial cosmology that we adopted for this analysis is the standard $\Lambda$CDM model detailed in the Introduction, having $f_\mathrm{NL} = 0$.

In this work we limited ourselves to study the level of non-Gaussianity $f_\mathrm{NL}$, and the least well-known of the standard cosmological parameters, $\sigma_8$. By the time Euclid will fly, it is reasonable to expect that all the other cosmological parameters will be known with great accuracy thanks to CMB experiments, such as \emph{Planck}, in combination with $H_0$ measurements. Hence it is fair to fix them to their fiducial values for our purposes. One note of caution with respect to the Fisher matrix analysis is in order. Here we assumed that the convergence power spectrum estimates obtained through shear, flexion, and their cross-correlation are independent of each other, so that the respective Fisher matrices can be simply summed. This assumption is acceptable for what concerns shear and flexion estimators, since they are measured from different moments of the galaxy light distributions. Still, when the cross-correlation between shear and flexion becomes relevant there might be other contributions to the Fisher matrix that have been neglected here \cite{VI12.1}. Further studies will be needed on this point.

\begin{figure}
\centering
\includegraphics[width=0.8\hsize]{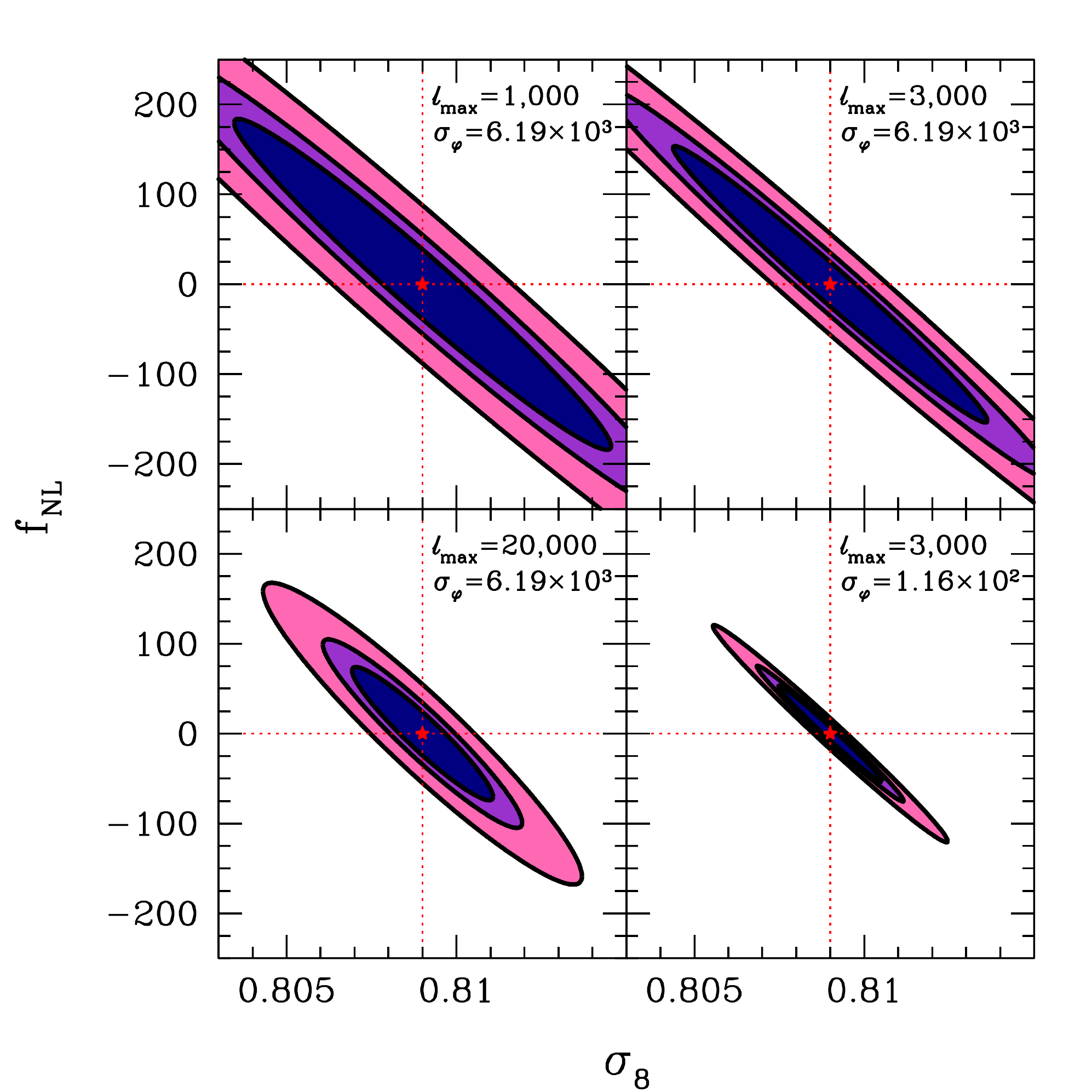}
\caption{Joint confidence regions for the parameters $\sigma_8$ and $f_\mathrm{NL}$ (local bispectrum shape) given by the combination of cosmic shear, cosmic flexion, and their cross-correlation. The blue regions refer to $68.3\%$ CL, the violet ones to $95.4\%$ CL, and the pink ones to $99.7\%$ CL. The primordial bispectrum shape is assumed to be local. The red stars at the intersection of the red dotted line show the fiducial Gaussian cosmology. In the different panels we show different combinations of the maximum multipole $\ell_\mathrm{max}$ over which Fisher matrices are computed, and the intrinsic shape noise for flexion measurements $\sigma_\varphi$, as labeled.}
\label{fig:fisher}
\end{figure}

In Figure \ref{fig:fisher} we show the joint constraints on the $\sigma_8 - f_\mathrm{NL}$ plane for different CLs, given by the combination of the shear-based estimate of the convergence power spectrum, the flexion-based estimate, and their cross-correlation. In Table \ref{tab:parameters} we report the $1\sigma$ marginalized errors on the two individual  parameters $f_\mathrm{NL}$ and $\sigma_8$ obtained by using the same data. The marginalized $1\sigma$ constraint for the $j-$th parameter can be found by inverting the Fisher matrix and then computing $(F^{-1}_{jj})^{1/2}$. While $\ell_\mathrm{min} = 50$ is left fixed, we adopted three different values for the maximum multipole $\ell_\mathrm{max}$. In the first configuration we limited ourselves to relatively small multipoles, $\ell_\mathrm{max} = 1,000$, where the behavior of the matter power spectrum is sufficiently well understood, either thanks to numerical simulations or perturbation theory. In the second case we selected $\ell_\mathrm{max} = 3,000$, which is about the upper limit within which the non-Gaussian part of the covariance (at least for the shear-based estimator of the convergence) can be neglected \cite{WH00.1,CO01.1,TA07.1,TA09.1}, as can the inner structure of dark matter halos and the impact of baryonic physics \cite{JI06.1,RU08.2,FE12.1}. Finally, in the third situation we pushed the sum in Eqs. (\ref{eqn:fisher_shear}), (\ref{eqn:fisher_flexion}), and (\ref{eqn:fisher_cross}) up to $\ell_\mathrm{max} = 20,000$. This corresponds to an angular scale of $\theta \sim 1$ arcmin, and hence equals at considering the optimal situation in which the entire field covered by \emph{Euclid} is tesselated in pixels of $\sim 1$ arcmin$^2$ and each is usable for weak lensing studies. Although this latter configuration has been used by some authors (even recently, see \cite{GI11.1}), it is quite extreme. A fruitful exploitation of such small scales would require detailed knowledge of the inner structure of dark matter halos, the impact of gas cooling, star formation, and AGN feedback, as well as the non-Gaussian part of the covariance. The physics of baryons, in particular, is expected to have a substantial impact at these small scales \cite{VA11.1,FE12.1}, possibly erasing part of the non-Gaussian signal. As a consequence, the results for this configuration must be taken as rough approximations of the real situation.

It should be noted that the multipole interval recommended by the \emph{Euclid Red Book} ranges up to $\ell_\mathrm{max} = 5,000$. In our analysis however we adopted as a fiducial configuration $\ell_\mathrm{max} = 3,000$ for two reasons: first, we wanted to be sure that the non-Gaussian parts of the covariances are negligible, an assumption that loses more and more accuracy the smaller is the angular scale considered. Second, this allows a more direct comparison with our previous results in \cite{FE10.1}, where the exact same multipole interval has been used.

\begin{table*}
  \caption{The $1\sigma$ marginalized errors on $\sigma_8$ and $f_\mathrm{NL}$ for the various estimators of the convergence power spectrum considered in this work and their overall combination. The primordial non-Gaussian bispectrum is assumed to have local shape, and different combinations of $\ell_\mathrm{max}$ and $\sigma_\varphi$ have been considered (see the text for details).} \label{tab:parameters}
  \begin{center}
  {\footnotesize
    \begin{tabular}{lcccccccc}
      \toprule[0.4mm]
      &\multicolumn{6}{c}{$\sigma_\varphi = 6.19\times 10^{3}$}&\multicolumn{2}{c}{$\sigma_\varphi = 1.16\times 10^{2}$}\\
      \midrule
      	&\multicolumn{2}{c}{$\ell_\mathrm{max}=1,000$}&\multicolumn{2}{c}{$\ell_\mathrm{max}=3,000$}&\multicolumn{2}{c}{$\ell_\mathrm{max}=20,000$}&\multicolumn{2}{c}{$\ell_\mathrm{max}=3,000$}\\
      \midrule
      	&$\Delta\sigma_8$&$\Delta f_\mathrm{NL}$&$\Delta\sigma_8$&$\Delta f_\mathrm{NL}$&$\Delta\sigma_8$&$\Delta f_\mathrm{NL}$&$\Delta\sigma_8$&$\Delta f_\mathrm{NL}$\\
      \midrule
      cosmic shear &$3.67\times 10^{-3}$&$122$&$3.07\times 10^{-3}$&$102$&$1.48\times 10^{-3}$&$52.6$&$3.07\times 10^{-3}$&$102$\\
      cosmic flexion &$3.07$&$9.48\times10^{4}$&$0.407$&$1.51\times 10^4$&$1.06\times 10^{-2}$&$782$&$1.28\times10^{-3}$&$45.3$\\
      cross-correlation &$3.96\times 10^{-2}$&$1.36\times10^3$&$2.96\times 10^{-2}$&$1.03\times 10^3 $&$4.13\times 10^{-3}$&$204$&$2.02\times 10^{-3}$&$69.3$\\
      total &$3.66\times 10^{-3}$&$121$&$3.05\times 10^{-3}$&$101$&$1.37 \times 10^{-3}$&$48.9$&$1.01\times 10^{-3}$&$35.2$\\
      \bottomrule[0.4mm]
    \end{tabular}
    }
  \end{center}
\end{table*}

\begin{figure}
\centering
\includegraphics[width=0.8\hsize]{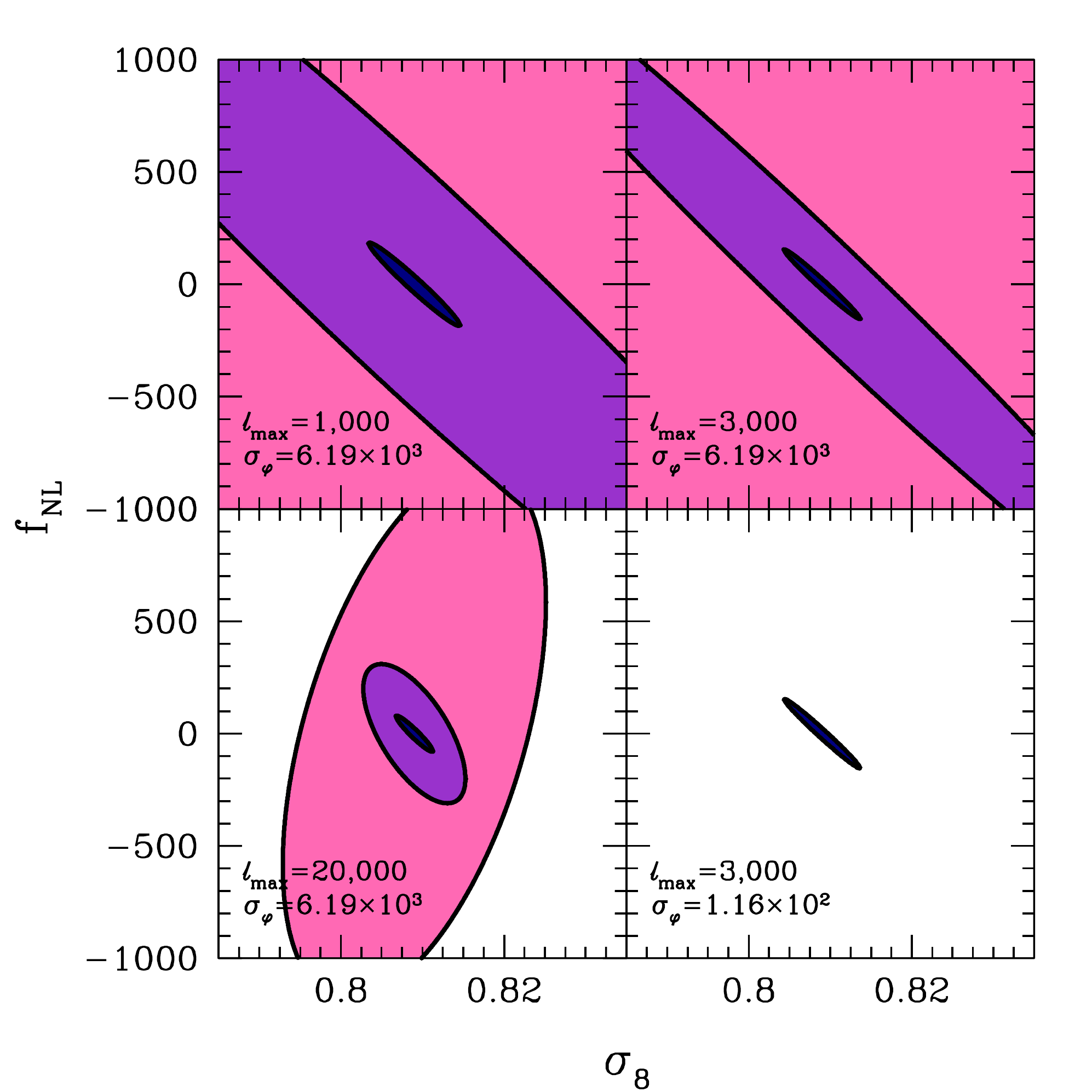}
\caption{The $68.3\%$ CL joint constraints on $\sigma_8$ and the level of PNG $f_\mathrm{NL}$ (local shape) given by cosmic flexion (pink contours), cosmic shear (blue), and their cross-correlation (violet). The shape of primordial bispectrum is assumed to be local. In the different panels we show different combinations of the maximum multipole $\ell_\mathrm{max}$ over which Fisher matrices are computed, and the intrinsic shape noise for flexion measurements $\sigma_\varphi$. Note that in the bottom right panel the contours stemming from flexion and the shear-flexion cross-correlation are not visible, being much smaller than the shear contour.}
\label{fig:fisher_combination}
\end{figure}

Figure \ref{fig:fisher} and Table \ref{tab:parameters} show that by using the cosmic shear alone it is possible to constrain $\sigma_8$ at the level of $\sim 3.7\times 10^{-3}$ and $f_\mathrm{NL}$ at the level of $\sim 120$ when $\ell_\mathrm{max} = 1,000$. These constraints are only mildly improved when considering smaller scales, being reduced by $\sim 20\%$ if $\ell_\mathrm{max} = 3,000$ and by a factor of $\sim 2$ if $\ell_\mathrm{max} = 20,000$. The constraints on $f_\mathrm{NL}$ are a factor of a few looser than those found in a previous work of ours, \cite{FE10.1}. The reason for this is that at the time the specifications for \emph{Euclid} were still not well defined, hence we adopted a background galaxy number density $\sim 33\%$ higher and an intrinsic shear noise $\sim 36\%$ smaller than in the present work. This resulted in a shot noise $\sim 2.5$ times smaller than here. Additionally, the assumed sky coverage was also one third larger there than here, and the adopted source redshift distributions are different.

The constraints on cosmological parameters coming from flexion alone are much weaker than those due to shear alone. For $\ell_\mathrm{max} = 1,000$ the forecasted error on $\sigma_8$ is of a few, and that on $f_\mathrm{NL}$ is $\sim 10^{5}$. Since we assumed that the number density of background sources usable for shape measurement be the same in the two cases, this difference is obviously due to the intrinsic shape noise, that for flexion is about $4$ orders of magnitude larger than for the shear. At the same time however, the cosmic shear shot noise is white, while the cosmic flexion shot noise decreases toward small angular scales. As a consequence, constraints on cosmological parameters from flexion alone are improved by a factor of $\sim 6-7$ for $\ell_\mathrm{max} = 3,000$ and by more than two orders of magnitude when $\ell_\mathrm{max} = 20,000$. This fact can be found in Figure \ref{fig:fisher_combination}, where we compare the $68.3\%$ CL joint constraints on $\sigma_8$ and $f_\mathrm{NL}$ derived individually from cosmic shear, cosmic flexion, and their cross-correlation. Still, the constraints coming from cosmic flexion alone remain about one order of magnitude worse than those coming from shear alone. In order to have comparable constraints it would be necessary to extend the sum in the Fisher matrices substantially above the multipole for which the two statistical errors are comparable, $\ell \sim 20,000$.

\begin{table*}
  \caption{The correlation coefficients between $\sigma_8$ and $f_\mathrm{NL}$ for the various convergence power spectrum estimators considered in this work and their full combination. A local shape for the PNG bispectrum is assumed. Different combinations of $\ell_\mathrm{max}$ and $\sigma_\varphi$ are considered (see the text for details).} \label{tab:correlation}
  \begin{center}
  {\footnotesize
    \begin{tabular}{lcccc}
      \toprule[0.4mm]
      &\multicolumn{3}{c}{$\sigma_\varphi = 6.19\times 10^{3}$}&\multicolumn{1}{c}{$\sigma_\varphi = 1.16\times 10^{2}$}\\
      \midrule
      	&$\ell_\mathrm{max}=1,000$&$\ell_\mathrm{max}=3,000$&$\ell_\mathrm{max}=20,000$&$\ell_\mathrm{max}=3,000$\\
      \midrule
      cosmic shear &$-0.978$&$-0.987$&$-0.952$&$-0.987$\\
      cosmic flexion &$-0.995$&$-0.992$&$0.492$&$-0.990$\\
      cross-correlation &$-0.962$&$-0.985$&$-0.647$&$-0.990$\\
      total &$-0.977$&$-0.987$&$-0.945$&$-0.990$\\
      \bottomrule[0.4mm]
    \end{tabular}
    }
  \end{center}
\end{table*}

The constraints on cosmological parameters given by the cross-correlation of shear and flexion are in between the two above. In particular, said constraints improve with decreasing scale faster than those due to shear alone but slower than those due to flexion alone. In the ideal configuration having $\ell_\mathrm{max} = 20,000$, the constraints are a factor of $\sim 2-3$ better than those coming from flexion, but still a factor of $\sim 3-4$ worse than those coming from shear. As a consequence of the above discussion, the constraints on cosmological parameters given by shear are basically unchanged by the inclusion of both the flexion and the cross-correlation between shear and flexion. The exception to this is given by the configuration with $\ell_\mathrm{max} = 20,000$, for which constraints are improved by $\sim 8\%$ upon inclusion of the flexion information.

The improvement on cosmological constraints given by the inclusion of flexion for $\ell_\mathrm{max} = 20,000$ is not only due to the decrement of the statistical noise of the latter with increasing multipole, but also to another factor that can be better appreciated by examining Table \ref{tab:correlation}. This table reports the correlation coefficients for the two cosmological parameters considered here, obtained by adopting shear, flexion, and their cross-correlation individually, as well as their full combination. In general, the correlation coefficient between the $j-$th and the $k-$th parameters is given by

\begin{equation}
r_{jk} \equiv \frac{F_{jk}^{-1}}{\sqrt{F_{jj}^{-1}F_{kk}^{-1}}}~.
\end{equation}
As can be seen, in almost all circumstances the correlation coefficient between $\sigma_8$ and $f_\mathrm{NL}$ is very close to $-1$, meaning that the two parameters at hand are almost perfectly anti-correlated with respect to the cosmological probes considered here. The correlation for cosmic flexion is slightly smaller than for cosmic shear, however the difference is only at the level of $1-2\%$. This fact is of course expected because both the shear and the flexion covariances (as well as the covariance of their cross-correlation) depend in the same way on cosmological parameters. An exception is given by the configuration with $\ell_\mathrm{max} = 20,000$. In this case the correlation for cosmic shear remains quite close to $-1$, while that for cosmic flexion changes sign and becomes positive, taking a value of $\sim 0.5$. The correlation coefficient for the cross-correlation also shows a substantial increase, although it remains on the negative side. These changes can be better appreciated by looking at the bottom left panel of Figure \ref{fig:fisher_combination}.

\begin{figure}
\centering
\includegraphics[width=0.8\hsize]{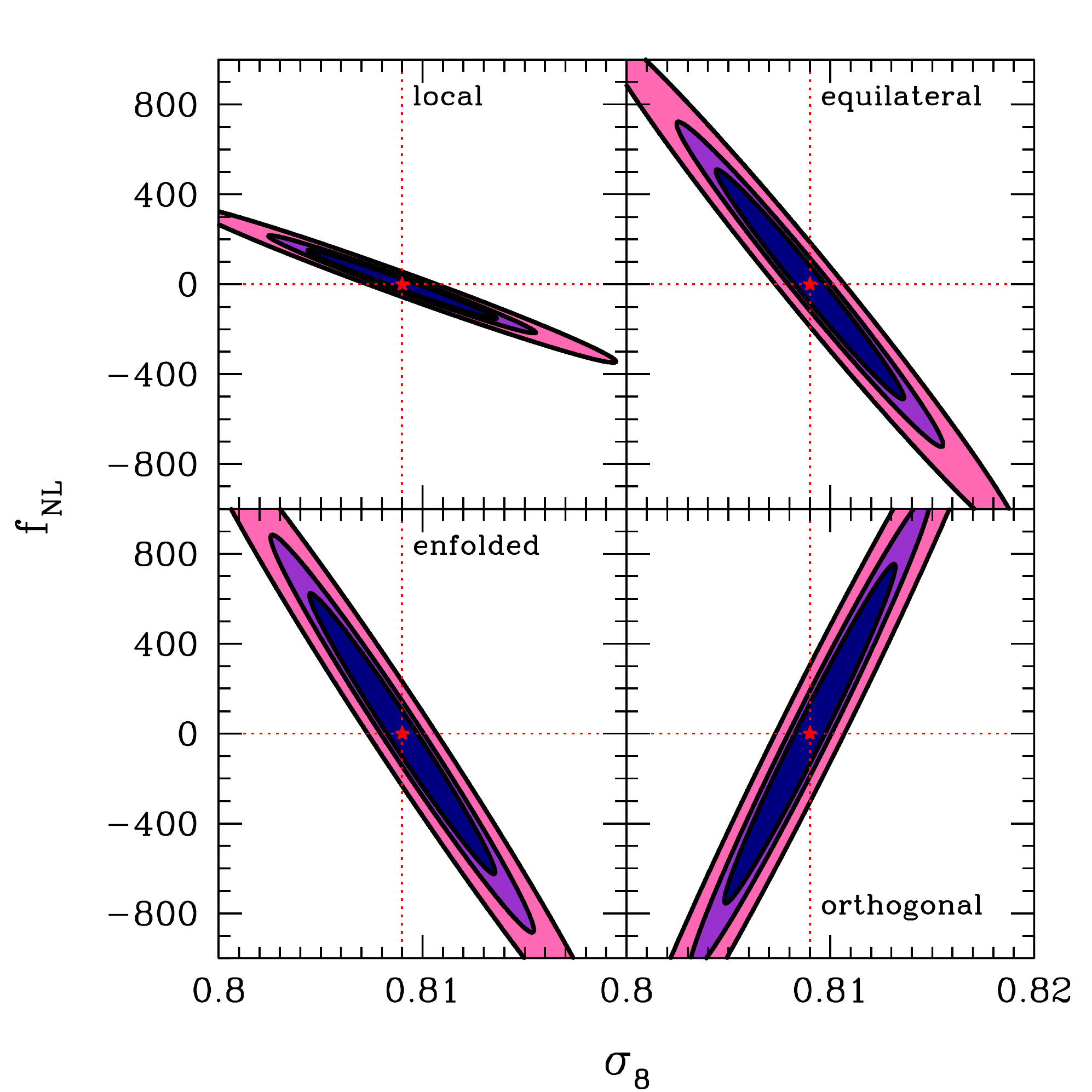}
\caption{Joint constraints on the level of non-Gaussianity $f_\mathrm{NL}$ and $\sigma_8$ given by the combination of cosmic shear, cosmic flexion, and their cross-correlation. Pink contours refer to $68.3\%$ CL, violet ones to $95.4\%$ CL, and blue ones to $99.7\%$ CL. Red stars at the intersections of red dotted lines mark the fiducial Gaussian cosmological model. Calculations assume $\ell_\mathrm{max} = 3,000$ and $\sigma_\varphi = 6.19\times 10^{3}$. Each panel refer to a different bispectrum shape, as labeled in the plot (see the text for more details).}
\label{fig:fisher_shape}
\end{figure}

The reason for this behavior is easily understood by looking at Figure \ref{fig:weakLensing}. At very small angular scales the effect of PNG on the convergence power spectrum gets reversed, with a positive $f_\mathrm{NL}$ value implying a reduction of power and vice-versa. Since both the estimators of the convergence power spectrum based on flexion and on the cross-correlation between shear and flexion are weighted toward high multipoles, they are more strongly affected by this behavior with respect to the shear-based estimator. It follows that the improvement in the constraints given by the inclusion of flexion over those of shear alone are partially due to the degeneracy-breaking impact of flexion. We stress again that this result should be taken with caution. Even ignoring the non-Gaussian part of the covariance, the improvement due to the inclusion of flexion could be either overestimated or underestimated: the fact that dark matter halos are more concentrated in non-Gaussian cosmologies with positive skewness would decrease the scale at which the impact of $f_\mathrm{NL}$ on the convergence power spectrum is reversed, hence reducing the improvement due to the inclusion of cosmic flexion. On the other hand, PNG might also imply a more intense formation of first stars \cite{MA11.1,MA11.2}, with the subsequent larger energy feedback which would effectively blow up the inner parts of dark matter halos and/or avoid gas condensation at the rates of a $\Lambda$CDM universe, and hence work in the opposite direction.

Let us now study in more detail the effect of the intrinsic flexion noise on the forecasting power of cosmological weak lensing. The value of $\sigma_\varphi$ that we adopted in the preceding discussion was considered standard by \cite{CA11.1}. At the same time, authors in \cite{PI10.1} adopted a value that is slightly larger, while other authors \cite{OK07.1,LE09.1} considered numbers that are a factor of $\sim 3$ smaller than that considered here. Despite the fact that the values of $\sigma_\varphi$ used in all these work share the same order of magnitude, it is worth stressing that a standard value for the flexion intrinsic noise does not exist yet. The values selected in those of the aforementioned works that attempt an actual measurement of $\sigma_\varphi$ both neglect the shear-flexion cross-talk \cite{VI12.1} and make use of non-optimized weight functions for flexion measurements, hence they cannot be considered as thoroughly reliable. More specifically, the first flexion that we are considering here induces arc-like distortions on the images of background galaxies. Yet, galaxies are intrinsically very rarely arc-shaped, so a value of $\sigma_\varphi$ as high as the one adopted above might be unlikely. While pixel noise and measurement noise can indeed introduce arc-shaped distortions in galaxy images, these are systematic effects that are beyond the scope of the present statistical analysis.

In order to estimate the impact of this uncertainty on cosmological constraints and to bracket viable alternatives, we repeated the Fisher matrix analysis for the configuration with $\ell_\mathrm{max} = 3,000$, by replacing the $\sigma_\varphi$ used above with a substantially lower value. We set this value by requiring that the flexion S/N matches the shear S/N at the smallest scale in the configuration at hand, that is $\ell = 3,000$. For a Singular Isothermal Sphere (SIS) the ratio between the flexion field and the shear field is $1/\theta$, where $\theta$ is the angular separation from the center of the sphere (see, e.g., \cite{LE10.1}). In the very outskirts of the SIS, say $\theta = 1$~arcmin, this ratio equals $1/(60~\mathrm{arcsec})$. For a typical image $1$~arcsec across, the ratio of the flexion signal to the shear signal is hence $1/60 = 0.0167$. Assuming that this holds also for LSS lensing and requiring that the flexion S/N and the shear S/N be the same at $\ell = 3,000$ implies $\sigma_\varphi \simeq 387\sigma_\gamma \simeq 116$. Although this choice admittedly bears some degree of arbitrariness, we feel it is the best that can be done until more detailed work on the topic at hand is performed. Moreover, this option also allows us to understand what level of $\sigma_\varphi$ would be necessary in order to obtain significant constraints on $f_\mathrm{NL}$ from cosmic flexion.

The results for the joint constraints on $\sigma_8$ and $f_\mathrm{NL}$ implied by the new value of the intrinsic flexion noise are reported in the last two columns of Table \ref{tab:parameters}, while the relative correlation coefficients are displayed in the last column of Table \ref{tab:correlation}. As can be seen by inspecting these values, constraints due to cosmic flexion alone are improved by more than two orders of magnitude, while those due to the shear-flexion cross-correlation are improved by more than one order of magnitude. More importantly, both constraints are now tighter than those given by cosmic shear alone, by up to a factor of $\sim 2$ for cosmic flexion. This result might look surprising at first, since the intrinsic shape noise for flexion is still more than two orders of magnitude larger than that for shear. However, due to the scale dependence of the flexion shot noise, the latter becomes smaller than the shear shot noise already at $\ell \gtrsim 400$, and since the sum extends up to $\ell = 3,000$ overall the flexion becomes the dominant contribution. As a result, the total combination of shear, flexion, and their cross-correlation improves constraints on PNG by a factor of $\sim 3$ with respect to the shear alone. 

The dramatic change in constraining power due to a reduction in the flexion intrinsic noise can be appreciated also by looking at the bottom right panels of Figures \ref{fig:fisher} and \ref{fig:fisher_combination}. In the former we show the joint constraints on $\sigma_8$ and $f_\mathrm{NL}$ given by the overall combination of the three estimators. The confidence ellipses are substantially smaller even than those resulting from the configuration with $\ell_\mathrm{max} = 20,000$ but with a high value of $\sigma_\varphi$. In the latter we compare the joint constraints given by each individual estimator taken separately. The difference in the sizes of the confidence ellipses with respect to the previously considered cases is so considerable that on the scale of the Figure only the shear constraints are visible, while the constraints stemming from flexion or the cross-correlation of shear and flexion are substantially smaller than those. Note the stark difference with the upper panels of the same Figure, where the confidence ellipse given by cosmic flexion alone did not even fit the scale.

As a final step, we considered again the configuration with $\ell_\mathrm{max} = 3,000$ and $\sigma_\varphi = 6.19\times 10^3$ and repeated the Fisher matrix analysis described above for three additional shapes of the PNG bispectrum, namely the equilateral, enfolded, and orthogonal shapes, as described in Section \ref{sct:ng}. The results for the joint constraints on $\sigma_8$ and $f_\mathrm{NL}$ obtained from the total combination of cosmic shear, cosmic flexion, and their cross-correlation are shown in Figure \ref{fig:fisher_shape}, while in Tables \ref{tab:parameters_shape} and \ref{tab:correlation_shape} we report the marginalized constraints for each individual probe and the correlation coefficients, respectively. As expected, forecasted errors on $\sigma_8$ are almost unchanged when considering different bispectrum shapes. On the other hand, constraints on the level of PNG are significantly looser for shapes different from the local one. This is so because the impact of non-Gaussianity on both the mass function and halo bias (and hence on the non-linear matter power spectrum) is weaker for these shapes (e.g., \cite{LO08.1,FE11.1}). Constraints on the level of PNG $f_\mathrm{NL}$ range from $\sim 340$ for the equilateral shape up to $\sim 500$ for the orthogonal shape, a factor of up to five larger than the constraints for the local shape.

\begin{table*}
  \caption{The $1\sigma$ marginalized errors on $\sigma_8$ and $f_\mathrm{NL}$ for the various convergence power spectrum estimators considered in this work and their overall combination. Four different bispectrum shapes are shown, as labeled. Calculations assume $\ell_\mathrm{max} = 3,000$ and $\sigma_\varphi = 6.19\times 10^3$.} \label{tab:parameters_shape}
  \begin{center}
  {\footnotesize
    \begin{tabular}{lcccccccc}
      \toprule[0.4mm]
      	&\multicolumn{2}{c}{local}&\multicolumn{2}{c}{equilateral}&\multicolumn{2}{c}{enfolded}&\multicolumn{2}{c}{orthogonal}\\
      \midrule
      	&$\Delta\sigma_8$&$\Delta f_\mathrm{NL}$&$\Delta\sigma_8$&$\Delta f_\mathrm{NL}$&$\Delta\sigma_8$&$\Delta f_\mathrm{NL}$&$\Delta\sigma_8$&$\Delta f_\mathrm{NL}$\\
      \midrule
      cosmic shear &$3.07\times 10^{-3}$&$102$&$3.06\times 10^{-3}$&$339$&$3.02\times 10^{-3}$&$414$&$2.78\times 10^{-3}$&$500$\\
      cosmic flexion &$0.407$&$1.51\times10^{4}$&$0.380$&$4.76\times 10^4$&$0.406$&$6.20\times 10^4$&$0.442$&$8.64\times 10^{4}$\\
      cross-correlation &$2.96\times 10^{-2}$&$1.03\times10^3$&$2.89\times 10^{-2}$&$3.37\times 10^3 $&$2.94\times 10^{-2}$&$4.22\times 10^3$&$2.89\times 10^{-2}$&$5.39\times 10^3$\\
      total &$3.05\times 10^{-3}$&$101$&$3.04\times 10^{-3}$&$337$&$3.00 \times 10^{-3}$&$412$&$2.77\times 10^{-3}$&$498$\\
      \bottomrule[0.4mm]
    \end{tabular}
    }
  \end{center}
\end{table*}

\begin{table*}
  \caption{The correlation coefficients between $\sigma_8$ and $f_\mathrm{NL}$ for the various convergence power spectrum estimators considered in this work and their overall combination. Four different bispectrum shapes for PNG are assumed, and calculations are performed with $\ell_\mathrm{max} = 3,000$ and $\sigma_\varphi = 6.19\times 10^3$.} \label{tab:correlation_shape}
  \begin{center}
  {\footnotesize
    \begin{tabular}{lcccc}
      \toprule[0.4mm]
      	&local&equilateral&enfolded&orthogonal\\
      \midrule
      cosmic shear &$-0.987$&$-0.987$&$-0.986$&$0.984$\\
      cosmic flexion &$-0.992$&$-0.990$&$-0.992$&$0.993$\\
      cross-correlation &$-0.985$&$-0.984$&$-0.985$&$0.984$\\
      total &$-0.987$&$-0.987$&$-0.987$&$0.984$\\
      \bottomrule[0.4mm]
    \end{tabular}
    }
  \end{center}
\end{table*}

Since we considered a high value of $\sigma_\varphi$, the inclusion of the flexion information does not bring any significant improvement on cosmological constraints for these non-Gaussian shapes, with the bounds on $f_\mathrm{NL}$ improving only by less than a percent upon combination of cosmic shear, cosmic flexion, and their cross-correlation. Also, the correlation coefficients are all very close to $-1$. The only exception is represented by the orthogonal model. In this model, the skewness of the density fluctuation distribution is well known to be positive for negative values of $f_\mathrm{NL}$ and vice-versa, so that the direction of degeneracy between $\sigma_8$ and $f_\mathrm{NL}$ is reversed with respect to other cases. This is very well visible in the bottom right panel of Figure \ref{fig:fisher_shape}. The forecasted constraints on the level of PNG obtained for bispectrum shapes other than the local one are roughly at the same level of the current bounds, derived either from the CMB or LSS tracers (see the discussion in Section \ref{sct:ng}). As explicitly shown for the local shape, a reduction of the flexion intrinsic shape noise would arguably improve these constraints by a substantial amount. In order to correctly address this issue however it is necessary to measure this intrinsic noise in a way that is both unbiased and optimized. Substantial work is thus still required in this direction.

%%%%%%%%%%%%%%%%%%%%%%%%%%%%%%%%%%%%%%%%%%%%%%%%%%%
\section{Discussion and conclusions}\label{sct:conclusions}
%%%%%%%%%%%%%%%%%%%%%%%%%%%%%%%%%%%%%%%%%%%%%%%%%%%

In this work we analyzed the constraints on PNG that will be put by future wide-field weak lensing surveys such as \emph{Euclid}. We considered shear, flexion, and their cross-correlation as estimators for the convergence power spectrum. In Appendix \ref{sct:appendix} we collected detailed calculations of the relevant covariances and Fisher matrices, which have been used in order to compute joint constraints for the level of PNG $f_\mathrm{NL}$ and the amplitude of the matter power spectrum $\sigma_8$ using the accepted specifications for \emph{Euclid} \cite{LA11.1}. Our main results can be summarized as follows.

\begin{itemize}
\item The latest \emph{Euclid} specifications imply somewhat smaller sky coverage and average galaxy number density than assumed in previous calculations. As a result, the constraining power of cosmic shear is weaker than formerly found. Specifically, the level of PNG for local shape bispectrum will be constrained at the level of $\Delta f_\mathrm{NL} \sim 100$, which is at the same level of present-day constraints from the CMB and the LSS \cite{SL08.1,XI11.1}.
\item For the first time in the literature we forecasted joint \emph{Euclid} constraints on $f_\mathrm{NL}$ and $\sigma_8$ from cosmic shear for four different shapes of the primordial bispectrum, namely local, equilateral, enfolded, and orthogonal. Constraints on the latter three non-Gaussian shapes are substantially looser than for the former, ranging between $\sim 340-500$, again at the level of current constraints from the CMB and the LSS.
\item We carefully reviewed the calculation of the covariance for a shear-based estimator of the convergence power spectrum, and used the same approach to compute, also for the first time, the covariances of a flexion-based estimator and of an estimator based on the cross-correlation of shear and flexion. These results have then been used to estimate the relevant Fisher matrices and hence combine the shear information with the flexion one.
\item The constraining power of cosmic flexion depends heavily on the assumed intrinsic shape noise $\sigma_\varphi$. Values adopted by several previous works might be too high to be realistic, and result in a constraining power for cosmic flexion alone that is much lower than for the cosmic shear: constraints on $\sigma_8$ are of order unity, while those on $f_\mathrm{NL}$ are of the order of a few $\times 10^4$. The shear-flexion cross correlation is in between the two, with constraints still being barely significant with respect to cosmic shear alone.
\item Since the shot noise for cosmic shear is white, while that for cosmic flexion and for their cross-correlation decreases with decreasing angular scale, the higher the multipole considered in the Fisher matrix analysis, the more the constraining power of flexion grows with respect to that of the shear. Pushing the sum up to $\ell_\mathrm{max} = 20,000$ improves the total joint constraints on $f_\mathrm{NL}$ and $\sigma_8$ by about $\sim 8\%$ with respect to the cosmic shear alone. Considering even larger multipoles would increase this relative contribution even more.
\item The value of $\sigma_\varphi$ suggested in the literature is likely neither unbiased nor optimal. A very rough order-of-magnitude estimate suggests a value for $\sigma_\varphi$ a factor of $\sim 50$ smaller than that, which produces an enormous improvement on the cosmological constraints due to flexion-related estimators. For local shape PNG flexion bounds become a factor of $\sim 2$ better than the shear bounds, and the overall combination of the three estimators performs a factor of $\sim 3$ better.
\end{itemize}

As mentioned above, constraints on PNG obtained by setting $\ell_\mathrm{max} = 3,000$ are comparable with the current bounds derived from the CMB and the LSS \cite{KO11.1,XI11.1}. This is true for all the four bispectrum shapes that have been considered in this work. Only pushing the sum in the Fisher matrix analysis up to $\ell_\mathrm{max} = 20,000$ can produce forecasted errors that are substantially lower than the current constraints. However, we stress once more that these results should be taken with extreme caution. Firstly because we ignored the non-Gaussian part of the covariance in our analysis, stemming from the non-linear mixing of modes, which might be important at very small scales. Secondly because the non-linear clustering of matter and the effect of baryonic physics at small scales are still very uncertain (for a recent discussion see \cite{VA12.1}). This consideration suggests several possible lines for future investigation.

The non-Gaussian contribution to the cosmic shear covariance has been estimated in \cite{TA09.1}, and using a similar approach it should be possible to do the same for cosmic flexion and for their cross-correlation. Substantial effort is already being put in order to gauge the impact of gas cooling, star formation, and AGN feedback on the small-scale clustering of matter \cite{RU08.2,VA11.1,FE12.1}. Although at the moment substantially scattered results are obtained due to the different implementations of non-gravitational physics, it is arguable that this issue will be much better under control by the time \emph{Euclid} data will be available. Last but not least, it is important to better understand the impact of PNG on the internal structure of dark matter halos. While it is relatively certain that halos in non-Gaussian cosmologies with positive skewness will be more compact than their Gaussian counterparts, there still are many obscure points. Just to mention one, nobody has yet investigated if indeed the NFW shape is  such a good fit to the density profiles of dark matter halos in models with PNG as it is for Gaussian cosmologies.

The results of this paper show that if the flexion intrinsic shape noise is at the level quoted by several previous works, cosmic flexion alone will not be usable for constraining cosmological parameters, not even with future state-of-the-art space missions, unless a better exploitment of highly non-linear scales becomes feasible. If $\sigma_\varphi$ turns out to be smaller instead, constraints due to cosmic flexion can be comparable or better than those due to cosmic shear. In the latter case, flexion will be a very important addition to the weak lensing analysis, and will allow to substantially improve constraints on PNG. In the former, the contribution of cosmic flexion to the constraining power of cosmological weak lensing can be safely neglected. Either way, it should be borne in mind that the final \emph{Euclid} data product will not be limited to weak lensing alone, and will overall allow to constrain the level of PNG at the level of $\Delta f_\mathrm{NL} \sim$ a few \cite{VE09.1}.

%%%%%%%%%%%%%%%%%%%%%%%%%%%%%%%%%%%%%%%%%%%%%%%%%%%
\section*{Acknowledgments}
%%%%%%%%%%%%%%%%%%%%%%%%%%%%%%%%%%%%%%%%%%%%%%%%%%%

CF is supported by the University of Florida through the Theoretical Astrophysics Fellowship. LM acknowledges financial contributions from contracts ASI-INAF I/023/05/0, ASI-INAF I/088/06/0, ASI I/016/07/0 'COFIS', ASI 'Euclid-DUNE' I/064/08/0, ASI-Uni Bologna-Astronomy Dept. 'Euclid-NIS' I/039/10/0, and PRIN MIUR 'Dark energy and cosmology with large galaxy surveys'. We are grateful to an anonymous referee for useful comments that allowed to improve significantly the presentation of this work.

\bibliographystyle{JHEP}
\bibliography{master}

\appendix
%%%%%%%%%%%%%%%%%%%%%%%%%%%%%%%%%%%%%%%%%%%%%%%%%%%
\section{Covariances and Fisher Matrices for Cosmological Weak Lensing}\label{sct:appendix}
%%%%%%%%%%%%%%%%%%%%%%%%%%%%%%%%%%%%%%%%%%%%%%%%%%%

In this Appendix we describe in detail how we computed the covariances of cosmic shear, cosmic flexion, and their cross-correlation, as well as the related Fisher matrices. First, let us recall how the shear is related to the convergence in Fourier space, namely

\begin{equation}
\hat\gamma(\boldsymbol\ell) = \frac{1}{\ell^2}\left( \ell_1^2-\ell_2^2 + 2\mathrm{i}\ell_1\ell_2 \right) \hat\kappa(\boldsymbol\ell).
\end{equation}
From this relation it easily follows that

\begin{equation}
\hat\gamma(\boldsymbol\ell)\hat\gamma^*(\boldsymbol\ell) = \hat\kappa(\boldsymbol\ell)\hat\kappa^*(\boldsymbol\ell)~,
\end{equation}
which motivates us to redefine the shear in Fourier space simply as $\hat\gamma(\boldsymbol\ell) = \hat\kappa(\boldsymbol\ell)$. In other words, we are adopting a scalar representation for the shear tensor field.

Similar considerations can be applied to the flexion. We shall limit ourselves to the first flexion $\varphi(\boldsymbol\theta)$, leaving the second flexion for future studies. First, let us recall the definition of the complex flexion, $\varphi(\boldsymbol\theta) = \partial \kappa(\boldsymbol\theta) = \partial_1\kappa(\boldsymbol\theta) + \mathrm{i}\partial_2\kappa(\boldsymbol\theta)$. Then, from the properties of the Fourier transform it easily follows that

\begin{equation}\label{eqn:flexion}
\hat\varphi(\boldsymbol\ell) = \hat\kappa(\boldsymbol\ell)\left(\mathrm{i} \ell_1 - \ell_2 \right)~,
\end{equation}
so that

\begin{equation}\label{eqn:mod}
\hat \varphi(\boldsymbol\ell)\hat \varphi^*(\boldsymbol\ell) = \hat\kappa(\boldsymbol\ell)\hat\kappa^*(\boldsymbol\ell)\left( \ell_1^2+\ell_2^2 \right) = \ell^2~\hat\kappa(\boldsymbol\ell)\hat\kappa^*(\boldsymbol\ell)~.
\end{equation}
Hence we redefine the flexion in Fourier space as $\hat\varphi(\boldsymbol\ell) = \ell~\hat\kappa(\boldsymbol\ell)$ (we might call this the \emph{absolute flexion}, as it loses the directional information). We recall that the convergence is a real-valued field, hence its Fourier transform enjoys the symmetry $\hat\kappa^*(\boldsymbol\ell) = \hat\kappa(-\boldsymbol\ell)$. The redefined shear and flexion also share the same symmetry, as it is easy to verify.

By using the Limber approximation the power spectrum of the convergence can be written as

\begin{equation}
\left\langle \hat{\kappa}(\boldsymbol\ell)\hat{\kappa}^*(\boldsymbol\ell') \right\rangle  = (2\pi)^2 \delta_\mathrm{D}(\boldsymbol\ell-\boldsymbol\ell') P_\kappa(\boldsymbol\ell)~,
\end{equation}
where the angular brackets represent the ensemble average and $P_\kappa(\boldsymbol\ell) = \ell^4P_\phi(\boldsymbol \ell)$, with $P_\phi(\boldsymbol\ell)$ being the power spectrum of the lensing potential $\phi$ (see also Eq. \ref{eqn:convergenceps}). By using the previous relations, the power spectrum of the flexion can now be written as

\begin{equation}
\left\langle \hat \varphi(\boldsymbol\ell)\hat \varphi^*(\boldsymbol\ell') \right\rangle  = (2\pi)^2 \delta_\mathrm{D}(\boldsymbol\ell-\boldsymbol\ell') P_\varphi(\boldsymbol\ell) = \ell\ell'(2\pi)^2 \delta_\mathrm{D}(\boldsymbol\ell-\boldsymbol\ell') P_\kappa(\boldsymbol\ell)~.
\end{equation}
It follows that the flexion power spectrum can be related to the convergence power spectrum by $P_\varphi(\boldsymbol\ell) = \ell^2P_\kappa(\boldsymbol\ell)$. This is a well known relation obtained for the first time by \cite{BA06.2}. Analogously, the cross-power spectrum between the convergence and the flexion reads

\begin{equation}
\frac{1}{2}\left[\left\langle \hat\kappa(\boldsymbol\ell)\hat \varphi^*(\boldsymbol\ell') \right\rangle + \left\langle \hat \varphi(\boldsymbol\ell)\hat\kappa^*(\boldsymbol\ell') \right\rangle\right] = (2\pi)^2 \delta_\mathrm{D}(\boldsymbol\ell-\boldsymbol\ell') P_x(\boldsymbol\ell) = \frac{1}{2}\left( \ell+\ell'\right) (2\pi)^2 \delta_\mathrm{D}(\boldsymbol\ell-\boldsymbol\ell') P_\kappa(\boldsymbol\ell)~,
\end{equation}
which implies that the cross correlation between the convergence and the flexion can be related to the convergence power spectrum as $P_x(\boldsymbol\ell) = \ell~P_\kappa(\boldsymbol\ell)$. This also coincides with the result shown by \cite{BA06.2}.

%%%%%%%%%%%%%%%%%%%%%%%%%%%%%%%%%%%%%%%%%%%%%%%%%%%
\subsection{Estimators}
%%%%%%%%%%%%%%%%%%%%%%%%%%%%%%%%%%%%%%%%%%%%%%%%%%%

In order to compute the covariances, we need first to quantify estimators for the convergence power spectrum as measured through shear and through flexion. For this we shall follow closely the recent work by \cite{PI10.1}. Let us suppose to have shape measurements (either ellipticity for the shear or some other moment of the luminosity distribution for the flexion) for a sample of $N$ galaxies uniformly distributed across a field of area $4\pi f_\mathrm{sky}$, so that the average number density of galaxies is $\bar n = N/(4\pi f_\mathrm{sky})$. According to the authors in \cite{KA98.1,JO08.1}, an estimator for the complex shear in Fourier space can be written as $\hat\gamma^\mathrm{(e)}(\boldsymbol\ell) = \bar n~\hat\gamma (\boldsymbol\ell) + \hat\varepsilon_\gamma(\boldsymbol\ell)$, where $\varepsilon_\gamma(\boldsymbol\theta)$ represents the intrinsic contribution to the measured ellipticity, and since this can be measured only at an actual galaxy position, its Fourier transform can be discretized according to

\begin{equation}
\hat\varepsilon_\gamma(\boldsymbol\ell) = \sum_{j=1}^N \varepsilon_\gamma(\boldsymbol \theta_j) \exp\left({-\mathrm{i}\boldsymbol \ell \cdot \boldsymbol \theta_j}\right)~.
\end{equation}
We further recall that the scalar representation of the shear reads $\hat\gamma(\boldsymbol\ell) = \hat\kappa(\boldsymbol\ell)$, hence an estimator for the convergence based on shear measurements can be written as 

\begin{equation}
\hat\kappa_\gamma(\boldsymbol\ell) = \hat\gamma^\mathrm{(e)}(\boldsymbol\ell) = \bar n~\hat\kappa(\boldsymbol\ell) +\hat\varepsilon_\gamma(\boldsymbol\ell)~.
\end{equation}
In a completely analogous way, we can come up with an estimator for the flexion, $\hat \varphi^\mathrm{(e)}(\boldsymbol\ell) = \bar n~\hat \varphi(\boldsymbol\ell) + \hat\varepsilon_\varphi(\boldsymbol\ell)$, where the function $\varepsilon_\varphi(\boldsymbol\theta)$ encapsulates this time the intrinsic contribution to the measured flexion. Furthermore, because of the way we redefined the flexion here, we know that $\hat \varphi(\boldsymbol\ell) = \ell~\hat\kappa(\boldsymbol\ell)$, hence an estimator for the convergence based on flexion measurements can be written as 

\begin{equation}
\hat\kappa_\varphi(\boldsymbol\ell) = \bar n~\hat\kappa(\boldsymbol\ell) + \frac{1}{\ell}\hat\varepsilon_\varphi(\boldsymbol\ell)~.
\end{equation}
Note that, since the intrinsic shear and flexion are assumed to be randomly distributed, then $\langle \varepsilon_\gamma(\boldsymbol \theta_j)\rangle = 0 = \langle \varepsilon_\varphi(\boldsymbol \theta_j)\rangle$ for all $j$.

We now introduce a shear-based estimator for the convergence power spectrum as \cite{JO08.1}

\begin{equation}\label{eqn:shear_based}
P^{(\gamma)}_\kappa(\boldsymbol\ell) = \frac{1}{4\pi f_\mathrm{sky}\bar n^2}\int_{\Omega_{\boldsymbol\ell}} \frac{\d^2\boldsymbol\xi}{m(\Omega_{\boldsymbol\ell})}\hat\kappa_\gamma(\boldsymbol\xi)\hat\kappa_\gamma^*(\boldsymbol\xi) - \frac{\sigma_\gamma^2}{\bar n}~,
\end{equation}
where $\Omega_{\boldsymbol\ell}$ is some small region in multipole space around $\boldsymbol\ell$, $m(\Omega_{\boldsymbol\ell})$ is its volume, and $f_\mathrm{sky}$ is the fraction of sky covered by the shear (flexion later on) measurement \cite{SC99.2}. The quantity $\sigma_\gamma$ represents the average dispersion of the intrinsic source ellipticity, that is defined as

\begin{equation}
\left\langle\varepsilon_\gamma (\boldsymbol\theta_j)\varepsilon_\gamma^* (\boldsymbol\theta_k) \right\rangle = \delta_\mathrm{K}^{jk} \sigma_\gamma^2~.
\end{equation} 
The ensemble average of this estimator reads

\begin{equation}
\left\langle P^{(\gamma)}_\kappa(\boldsymbol\ell)\right\rangle =  \frac{1}{4\pi f_\mathrm{sky}\bar n^2}\int_{\Omega_{\boldsymbol\ell}} \frac{\d^2\boldsymbol\xi}{m(\Omega_{\boldsymbol\ell})} \left\langle \hat\kappa_\gamma(\boldsymbol\xi)\hat\kappa_\gamma^*(\boldsymbol\xi) \right\rangle -\frac{\sigma_\gamma^2}{\bar n}~,
\end{equation}
where the argument of the integral in the previous equation can be estimated thanks to the following reasoning,

\begin{equation}\label{eqn:arg}
\left\langle \hat\kappa_\gamma(\boldsymbol\ell)\hat\kappa^*_\gamma(\boldsymbol\ell') \right\rangle = \bar n^2(2\pi)^2\delta_\mathrm{D}(\boldsymbol\ell - \boldsymbol\ell') P_\kappa(\boldsymbol\ell) + \bar n \left\langle \hat\varepsilon_\gamma(\boldsymbol\ell)  \hat\kappa^*(\boldsymbol\ell') \right\rangle + \bar n\left\langle \hat\kappa(\boldsymbol\ell)\hat\varepsilon_\gamma^*(\boldsymbol\ell') \right\rangle + \left\langle \hat\varepsilon_\gamma(\boldsymbol\ell) \hat\varepsilon_\gamma^*(\boldsymbol\ell') \right\rangle~.
\end{equation}
The second and the third terms of the previous sum represent the cross-correlation between the convergence field and the intrinsic ellipticity of sources. The two can be considered to be statistically independent, so that these two terms vanish. The last term is instead the power spectrum of intrinsic ellipticity, and it is a little bit trickier to determine. Specifically,

\begin{eqnarray}
\left\langle \hat\varepsilon_\gamma(\boldsymbol\ell) \hat\varepsilon_\gamma^*(\boldsymbol\ell') \right\rangle &=& \left\langle \left[ \sum_{j=1}^N \varepsilon_\gamma(\boldsymbol \theta_j) \exp\left({-\mathrm{i}\boldsymbol \ell \cdot \boldsymbol \theta_j}\right) \right]\left[  \sum_{k=1}^N \varepsilon_\gamma^*(\boldsymbol \theta_k) \exp\left({\mathrm{i}\boldsymbol \ell' \cdot \boldsymbol \theta_k}\right) \right] \right\rangle = 
\nonumber\\
&=& \sum_{jk=1}^N\left\langle\varepsilon_\gamma (\boldsymbol\theta_j)\varepsilon_\gamma^* (\boldsymbol\theta_k) \right\rangle\exp\left( -\mathrm{i}\boldsymbol\ell \cdot \boldsymbol\theta_j +\mathrm{i} \boldsymbol\ell'\cdot\boldsymbol\theta_k  \right) = 
\nonumber\\
&=& \sigma_\gamma^2 \sum_{j=1}^N\exp\left[ -\mathrm{i}(\boldsymbol\ell-\boldsymbol\ell')\cdot \boldsymbol\theta_j\right] = \sigma^2_\gamma~\bar n~(2\pi)^2\delta_\mathrm{D}(\boldsymbol\ell - \boldsymbol\ell')~.
\end{eqnarray}
By replacing this result in the previous Eq. (\ref{eqn:arg}) we obtain for the ensemble average of the shear-based estimator of the convergence power spectrum

\begin{equation}
\left\langle P^{(\gamma)}_\kappa(\boldsymbol\ell)\right\rangle =  \frac{1}{4\pi f_\mathrm{sky}}\int_{\Omega_{\boldsymbol\ell}} \frac{\d^2\boldsymbol\xi}{m(\Omega_{\boldsymbol\ell})} (2\pi)^2\delta_\mathrm{D}(0)\left[ P_\kappa(\boldsymbol\xi) +\frac{\sigma_\gamma^2}{\bar n}\right] -\frac{\sigma_\gamma^2}{\bar n}~.
\end{equation}
By assuming that the region $\Omega_{\boldsymbol\ell}$ is small enough, so that the function in square brackets does not vary significantly across it, we can pull the function itself outside the integral. Moreover, the quantity $(2\pi)^2\delta_\mathrm{D}(\boldsymbol\ell)$ can be interpreted as the Fourier transform of the window function for the survey at hand, as long as the average separation between neighboring galaxies is much smaller than the extent of the field \cite{JO08.1}. When computed at zero, this just equals the area of the survey, $4\pi f_\mathrm{sky}$. It follows that 

\begin{equation}
\left\langle P^{(\gamma)}_\kappa(\boldsymbol\ell)\right\rangle \simeq P_\kappa(\boldsymbol\ell)~.
\end{equation}
Hence the estimator for the convergence power spectrum defined in Eq. (\ref{eqn:shear_based}) is unbiased.

In a similar way, it is possible to define a flexion-based estimator for the convergence power spectrum, as

\begin{equation}\label{eqn:flexion_based}
P^{(\varphi)}_\kappa(\boldsymbol\ell) = \frac{1}{4\pi f_\mathrm{sky}\bar n^2}\int_{\Omega_{\boldsymbol\ell}} \frac{\d^2\boldsymbol\xi}{m(\Omega_{\boldsymbol\ell})}\hat\kappa_\varphi(\boldsymbol\xi)\hat\kappa_\varphi^*(\boldsymbol\xi) - \frac{\sigma_\varphi^2}{\bar n\ell^2}~,
\end{equation}
where in this case the average dispersion of the intrinsic flexion is defined by

\begin{equation}
\left\langle\varepsilon_\varphi (\boldsymbol\theta_j)\varepsilon_\varphi^* (\boldsymbol\theta_k) \right\rangle = \delta_\mathrm{K}^{jk} \sigma_\varphi^2~.
\end{equation} 
Note that in this case the last term on the right-hand side of Eq. (\ref{eqn:flexion_based}) depends on the multipole, consequence of the fact that the noise for a flexion-based convergence estimate is not white \cite{PI10.1}. Specifically, by considering the ensemble average of the estimator in Eq. (\ref{eqn:flexion_based}), we obtain

\begin{equation}
\left\langle P^{(\varphi)}_\kappa(\boldsymbol\ell)\right\rangle =  \frac{1}{4\pi f_\mathrm{sky}}\int_{\Omega_{\boldsymbol\ell}} \frac{\d^2\boldsymbol\xi}{m(\Omega_{\boldsymbol\ell})} (2\pi)^2\delta_\mathrm{D}(0)\left[P_\kappa(\boldsymbol\xi) +\frac{\sigma_\varphi^2}{\bar n\xi^2}\right]-\frac{\sigma_\varphi^2}{\bar n\ell^2}~,
\end{equation}
and with the same approximation used above about $\Omega_{\boldsymbol\ell}$ it follows that

\begin{equation}
\left\langle P^{(\varphi)}_\kappa(\boldsymbol\ell)\right\rangle \simeq P_\kappa(\boldsymbol\ell)~.
\end{equation}

Similar considerations can be applied in order to construct an estimator of the convergence power spectrum that is based on the cross-correlation of shear and flexion. Such an estimator can be written as

\begin{equation}\label{eqn:both_based}
P_\kappa^{(x)} (\boldsymbol\ell) = \frac{1}{2}\frac{1}{4\pi f_\mathrm{sky}\bar n^2}\int_{\Omega_{\boldsymbol\ell}} \frac{\d^2\boldsymbol\xi}{m(\Omega_{\boldsymbol\ell})} \left[ \hat\kappa_\gamma(\boldsymbol\xi)\hat\kappa^*_\varphi(\boldsymbol\xi) + \hat\kappa_\varphi(\boldsymbol\xi)\hat\kappa^*_\gamma(\boldsymbol\xi) \right] - \frac{\sigma^2_x}{\bar n\ell}~,
\end{equation}
where now

\begin{equation}
\left\langle\varepsilon_\gamma (\boldsymbol\theta_j)\varepsilon_\varphi^* (\boldsymbol\theta_k) \right\rangle = \left\langle\varepsilon_\varphi (\boldsymbol\theta_j)\varepsilon_\gamma^* (\boldsymbol\theta_k) \right\rangle = \delta_\mathrm{K}^{jk} \sigma_x^2~.
\end{equation}
The ensemble average of this estimator reads

\begin{equation}
\left\langle P_\kappa^{(x)} (\boldsymbol\ell)\right\rangle = \frac{1}{2}\frac{1}{4\pi f_\mathrm{sky}\bar n^2}\int_{\Omega_{\boldsymbol\ell}} \frac{\d^2\boldsymbol\xi}{m(\Omega_{\boldsymbol\ell})} \left[ \left\langle\hat\kappa_\gamma(\boldsymbol\xi)\hat\kappa^*_\varphi(\boldsymbol\xi)\right\rangle + \left\langle\hat\kappa_\varphi(\boldsymbol\xi)\hat\kappa^*_\gamma(\boldsymbol\xi)\right\rangle \right] - \frac{\sigma^2_x}{\bar n\ell}~.
\end{equation}
and adopting the same procedure highlighted above the argument of the integral in Eq. (\ref{eqn:both_based}) can be explicitly recast by using the following expression,

\begin{equation}
\left\langle \hat\kappa_\gamma(\boldsymbol\ell)\hat\kappa^*_\varphi(\boldsymbol\ell') \right\rangle + \left\langle \hat\kappa_\varphi(\boldsymbol\ell)\hat\kappa^*_\gamma(\boldsymbol\ell') \right\rangle = 2 (2\pi)^2\delta_\mathrm{D}(\boldsymbol\ell - \boldsymbol\ell') \left[ \bar n^2 P_\kappa(\boldsymbol\ell) + \bar n\frac{\sigma^2_x}{2} \left( \frac{1}{\ell} + \frac{1}{\ell'} \right) \right] 
\end{equation}
The last term of the previous equation simplifies when $\boldsymbol\ell = \boldsymbol\ell'$, so that

\begin{equation}
\left\langle P_\kappa^{(x)} (\boldsymbol\ell)\right\rangle = \frac{1}{4\pi f_\mathrm{sky}}\int_{\Omega_{\boldsymbol\ell}} \frac{\d^2\boldsymbol\xi}{m(\Omega_{\boldsymbol\ell})}(2\pi^2)\delta_\mathrm{D}(0) \left[ P_\kappa(\boldsymbol\xi) + \frac{\sigma_x^2}{\bar n\xi}\right] - \frac{\sigma^2_x}{\bar n\ell} \simeq P_\kappa(\boldsymbol\ell)~.
\end{equation}
Once again, the convergence power spectrum estimate resulting from the cross-correlation of shear and flexion has a colored noise, which decreases with increasing multipole, albeit more slowly than the noise on the purely flexion-based estimate.

%%%%%%%%%%%%%%%%%%%%%%%%%%%%%%%%%%%%%%%%%%%%%%%%%%%
\subsection{Covariances}
%%%%%%%%%%%%%%%%%%%%%%%%%%%%%%%%%%%%%%%%%%%%%%%%%%%

From the discussion above, it is quite clear that the covariance of the convergence power spectrum will depend on the observable that is adopted for estimating it. Let us start with the covariance of the shear-based estimate, 

\begin{equation}\label{eqn:cov_shear}
\Psi\left[ P_\kappa^{(\gamma)}(\boldsymbol\ell), P_\kappa^{(\gamma)}(\boldsymbol\ell') \right] = \left\langle P_\kappa^{(\gamma)}(\boldsymbol\ell) P_\kappa^{(\gamma)}(\boldsymbol\ell') \right\rangle - \left\langle P_\kappa^{(\gamma)}(\boldsymbol\ell)  \right\rangle\left\langle P_\kappa^{(\gamma)}(\boldsymbol\ell') \right\rangle \simeq \left\langle P_\kappa^{(\gamma)}(\boldsymbol\ell) P_\kappa^{(\gamma)}(\boldsymbol\ell') \right\rangle - P_\kappa(\boldsymbol\ell) P_\kappa(\boldsymbol\ell')~.
\end{equation}
The first term of the sum above requires knowledge of the convergence four point correlation function. Specifically,

\begin{eqnarray}
\left\langle P_\kappa^{(\gamma)}(\boldsymbol\ell) P_\kappa^{(\gamma)}(\boldsymbol\ell') \right\rangle &=& \frac{1}{(4\pi f_\mathrm{sky})^2\bar n^4}\int_{\Omega_{\boldsymbol\ell}}\frac{\d^2 \boldsymbol\xi}{m(\Omega_{\boldsymbol\ell})}\int_{\Omega_{\boldsymbol\ell'}}\frac{\d^2 \boldsymbol\xi'}{m(\Omega_{\boldsymbol\ell'})} \left\langle \hat\kappa_\gamma(\boldsymbol\xi) \hat\kappa_\gamma^*(\boldsymbol\xi) \hat\kappa_\gamma(\boldsymbol\xi') \hat\kappa_\gamma^*(\boldsymbol\xi')  \right\rangle - 
\nonumber\\
&-&\frac{\sigma_\gamma^2}{4\pi f_\mathrm{sky}\bar n^3}\int_{\Omega_{\boldsymbol\ell}}\frac{\d^2 \boldsymbol\xi}{m(\Omega_{\boldsymbol\ell})} \left\langle \hat\kappa_\gamma(\boldsymbol\xi) \hat\kappa_\gamma^*(\boldsymbol\xi) \right\rangle -
\nonumber\\
&-&\frac{\sigma_\gamma^2}{4\pi f_\mathrm{sky}\bar n^3}\int_{\Omega_{\boldsymbol\ell'}}\frac{\d^2 \boldsymbol\xi'}{m(\Omega_{\boldsymbol\ell'})} \left\langle \hat\kappa_\gamma(\boldsymbol\xi') \hat\kappa_\gamma^*(\boldsymbol\xi') \right\rangle + \frac{\sigma_\gamma^4}{\bar n^2}~.
\end{eqnarray}
According to the calculations performed in the previous Subsection, the integrals in the second and third terms on the right-hand side of the previous equation can be computed similarly to the ensemble average of $P_\kappa^{(\gamma)}(\boldsymbol\ell)$ and $P_\kappa^{(\gamma)}(\boldsymbol\ell')$, respectively, thus giving

\begin{eqnarray}\label{eqn:cov_shear2}
\left\langle P_\kappa^{(\gamma)}(\boldsymbol\ell) P_\kappa^{(\gamma)}(\boldsymbol\ell') \right\rangle &=& \frac{1}{(4\pi f_\mathrm{sky})^2\bar n^4}\int_{\Omega_{\boldsymbol\ell}}\frac{\d^2 \boldsymbol\xi}{m(\Omega_{\boldsymbol\ell})}\int_{\Omega_{\boldsymbol\ell'}}\frac{\d^2 \boldsymbol\xi'}{m(\Omega_{\boldsymbol\ell'})} \left\langle \hat\kappa_\gamma(\boldsymbol\xi) \hat\kappa_\gamma^*(\boldsymbol\xi) \hat\kappa_\gamma(\boldsymbol\xi') \hat\kappa_\gamma^*(\boldsymbol\xi')  \right\rangle - 
\nonumber\\
&-&\frac{\sigma_\gamma^2}{\bar n} \left[ P_\kappa(\boldsymbol\ell) + \frac{\sigma_\gamma^2}{\bar n} \right] - \frac{\sigma_\gamma^2}{\bar n} \left[ P_\kappa(\boldsymbol\ell') + \frac{\sigma_\gamma^2}{\bar n} \right] + \frac{\sigma_\gamma^4}{\bar n^2}~.
\end{eqnarray}

It is well established \cite{BE02.2,TA07.1} that thanks to the Wick's theorem \cite{WI50.1} the four point correlation function can be written as a combination of two point correlation functions, with an additional connected part remaining. In the specific case,

\begin{eqnarray}
\left\langle \hat\kappa_\gamma(\boldsymbol\xi) \hat\kappa_\gamma^*(\boldsymbol\xi) \hat\kappa_\gamma(\boldsymbol\xi') \hat\kappa_\gamma^*(\boldsymbol\xi')  \right\rangle &=& 
\left\langle \hat\kappa_\gamma(\boldsymbol\xi) \hat\kappa_\gamma^*(\boldsymbol\xi) \right\rangle \left\langle \hat\kappa_\gamma(\boldsymbol\xi') \hat\kappa_\gamma^*(\boldsymbol\xi') \right\rangle + 
\left\langle \hat\kappa_\gamma(\boldsymbol\xi) \hat\kappa_\gamma(\boldsymbol\xi') \right\rangle \left\langle \hat\kappa_\gamma^*(\boldsymbol\xi) \hat\kappa_\gamma^*(\boldsymbol\xi') \right\rangle +
\nonumber\\
&+&\left\langle \hat\kappa_\gamma(\boldsymbol\xi) \hat\kappa_\gamma^*(\boldsymbol\xi') \right\rangle \left\langle \hat\kappa_\gamma^*(\boldsymbol\xi) \hat\kappa_\gamma(\boldsymbol\xi') \right\rangle +
\left\langle \hat\kappa_\gamma(\boldsymbol\xi) \hat\kappa_\gamma^*(\boldsymbol\xi) \hat\kappa_\gamma(\boldsymbol\xi') \hat\kappa_\gamma^*(\boldsymbol\xi')  \right\rangle_\mathrm{c}~.
\end{eqnarray}
The double integral of the first term of the sum above gives

\begin{equation}
\frac{1}{(4\pi f_\mathrm{sky})^2\bar n^4}\int_{\Omega_{\boldsymbol\ell}}\frac{\d^2 \boldsymbol\xi}{m(\Omega_{\boldsymbol\ell})}\int_{\Omega_{\boldsymbol\ell'}}\frac{\d^2 \boldsymbol\xi'}{m(\Omega_{\boldsymbol\ell'})} \left\langle \hat\kappa_\gamma(\boldsymbol\xi) \hat\kappa_\gamma^*(\boldsymbol\xi) \right\rangle \left\langle \hat\kappa_\gamma(\boldsymbol\xi') \hat\kappa_\gamma^*(\boldsymbol\xi') \right\rangle = P_{\kappa\kappa}(\boldsymbol\ell)P_{\kappa\kappa}(\boldsymbol\ell') + \frac{\sigma_\gamma^2}{\bar n} P_{\kappa\kappa}(\boldsymbol\ell) + \frac{\sigma_\gamma^2}{\bar n} P_{\kappa\kappa}(\boldsymbol\ell') + \frac{\sigma_\gamma^4}{\bar n^2}~,
\end{equation}
which cancel, respectively, with the second term of the covariance (last term in Eq. \ref{eqn:cov_shear}) and with the last three terms of Eq. (\ref{eqn:cov_shear2}). There remain three terms in the covariance, namely

\begin{eqnarray}
\Psi\left[ P_{\kappa\kappa}^{(\gamma)}(\boldsymbol\ell), P_{\kappa\kappa}^{(\gamma)}(\boldsymbol\ell') \right] &=&
\frac{1}{(4\pi f_\mathrm{sky})^2\bar n^4}\int_{\Omega_{\boldsymbol\ell}}\frac{\d^2 \boldsymbol\xi}{m(\Omega_{\boldsymbol\ell})}\int_{\Omega_{\boldsymbol\ell'}}\frac{\d^2 \boldsymbol\xi'}{m(\Omega_{\boldsymbol\ell'})}\left\langle \hat\kappa_\gamma(\boldsymbol\xi) \hat\kappa_\gamma(\boldsymbol\xi') \right\rangle \left\langle \hat\kappa_\gamma^*(\boldsymbol\xi) \hat\kappa_\gamma^*(\boldsymbol\xi') \right\rangle +
\nonumber\\
&+&\frac{1}{(4\pi f_\mathrm{sky})^2\bar n^4}\int_{\Omega_{\boldsymbol\ell}}\frac{\d^2 \boldsymbol\xi}{m(\Omega_{\boldsymbol\ell})}\int_{\Omega_{\boldsymbol\ell'}}\frac{\d^2 \boldsymbol\xi'}{m(\Omega_{\boldsymbol\ell'})}\left\langle \hat\kappa_\gamma(\boldsymbol\xi) \hat\kappa_\gamma^*(\boldsymbol\xi') \right\rangle \left\langle \hat\kappa_\gamma^*(\boldsymbol\xi) \hat\kappa_\gamma(\boldsymbol\xi') \right\rangle +
\nonumber\\
&+&\frac{1}{(4\pi f_\mathrm{sky})^2\bar n^4}\int_{\Omega_{\boldsymbol\ell}}\frac{\d^2 \boldsymbol\xi}{m(\Omega_{\boldsymbol\ell})}\int_{\Omega_{\boldsymbol\ell'}}\frac{\d^2 \boldsymbol\xi'}{m(\Omega_{\boldsymbol\ell'})}\left\langle \hat\kappa_\gamma(\boldsymbol\xi) \hat\kappa_\gamma^*(\boldsymbol\xi) \hat\kappa_\gamma(\boldsymbol\xi') \hat\kappa_\gamma^*(\boldsymbol\xi')  \right\rangle_\mathrm{c}~.
\end{eqnarray}
The last term in the above sum is just an integral over the trispectrum of the estimated convergence, and arises due to the mode-coupling induced by the non-linear gravitational growth of structures. If one limits himself or herself to relatively large scales, the impact of this term can be neglected. We adopted this approximation in the present work, hence we simply label this integral as $\tau_\gamma(\boldsymbol\ell,\boldsymbol\ell')$ and disregard it later on. In order to make progress, we recall that $\hat\kappa(\boldsymbol\ell) = \hat\kappa^*(-\boldsymbol\ell)$. Moreover we assume that the real and imaginary parts of $\hat\varepsilon_\gamma(\boldsymbol\ell)$ are statistically independent \cite{JO08.1}. It easily follows that 

\begin{equation}
\left\langle \hat\varepsilon_\gamma(\boldsymbol\ell)\hat\varepsilon_\gamma(\boldsymbol\ell')  \right\rangle = \left\langle \hat\varepsilon_\gamma(\boldsymbol\ell)\hat\varepsilon_\gamma^*(-\boldsymbol\ell')  \right\rangle
\end{equation}
(despite being $\varepsilon_\gamma(\boldsymbol\theta)$ still a complex-valued function), and eventually

\begin{equation}
\left\langle \hat\kappa_\gamma(\boldsymbol\ell)\hat\kappa_\gamma(\boldsymbol\ell')  \right\rangle = \left\langle \hat\kappa_\gamma(\boldsymbol\ell)\hat\kappa_\gamma^*(-\boldsymbol\ell')  \right\rangle~.
\end{equation}

Making use of the former assumption we can recast the covariance of shear-based estimates as

\begin{eqnarray}
\Psi\left[ P_\kappa^{(\gamma)}(\boldsymbol\ell), P_\kappa^{(\gamma)}(\boldsymbol\ell') \right] &=&
\frac{1}{(4\pi f_\mathrm{sky})^2}\int_{\Omega_{\boldsymbol\ell}}\frac{\d^2 \boldsymbol\xi}{m(\Omega_{\boldsymbol\ell})}\int_{\Omega_{\boldsymbol\ell'}}\frac{\d^2 \boldsymbol\xi'}{m(\Omega_{\boldsymbol\ell'})} (2\pi)^2 \delta_\mathrm{D}(\boldsymbol\xi + \boldsymbol\xi') C_\kappa^{(\gamma)}(\boldsymbol\xi,\boldsymbol\xi') (2\pi)^2 \delta_\mathrm{D}(\boldsymbol\xi + \boldsymbol\xi') C_\kappa^{(\gamma)}(-\boldsymbol\xi',\boldsymbol\xi)+
\nonumber\\
&+&\frac{1}{(4\pi f_\mathrm{sky})^2}\int_{\Omega_{\boldsymbol\ell}}\frac{\d^2 \boldsymbol\xi}{m(\Omega_{\boldsymbol\ell})}\int_{\Omega_{\boldsymbol\ell'}}\frac{\d^2 \boldsymbol\xi'}{m(\Omega_{\boldsymbol\ell'})}(2\pi)^2 \delta_\mathrm{D}(\boldsymbol\xi - \boldsymbol\xi') C_\kappa^{(\gamma)}(\boldsymbol\xi,\boldsymbol\xi') (2\pi)^2 \delta_\mathrm{D}(\boldsymbol\xi - \boldsymbol\xi') C_\kappa^{(\gamma)}(\boldsymbol\xi',\boldsymbol\xi)+
\nonumber\\
&+&\tau_\gamma(\boldsymbol\ell,\boldsymbol\ell')~,
\end{eqnarray}
where

\begin{equation}
C_\kappa^{(\gamma)}(\boldsymbol\ell,\boldsymbol\ell') \equiv P_\kappa(\boldsymbol\ell) + \frac{\sigma_\gamma^2}{\bar n}~.
\end{equation}
Note that the function $C_\kappa^{(\gamma)}(\boldsymbol\ell,\boldsymbol\ell')$ is actually independent of the second argument, however we left it explicitly indicated for consistency with the flexion-based estimate discussed below. Now, since the convergence power spectrum depends only on the modulus of the multipole vector, it follows that $C_\kappa^{(\gamma)}(\boldsymbol\ell,\boldsymbol\ell') = C_\kappa^{(\gamma)}(-\boldsymbol\ell,\boldsymbol\ell')$. Also, we assume that the integration area $\Omega_{\boldsymbol\ell}$ is left unchanged under the transformation $\boldsymbol\ell \rightarrow - \boldsymbol\ell$, so that the previous equation can be recast as

\begin{equation}\label{eqn:cov_shear3}
\Psi\left[ P_\kappa^{(\gamma)}(\boldsymbol\ell), P_\kappa^{(\gamma)}(\boldsymbol\ell') \right] =
\frac{2}{(4\pi f_\mathrm{sky})^2}\int_{\Omega_{\boldsymbol\ell}}\frac{\d^2 \boldsymbol\xi}{m(\Omega_{\boldsymbol\ell})}\int_{\Omega_{\boldsymbol\ell'}}\frac{\d^2 \boldsymbol\xi'}{m(\Omega_{\boldsymbol\ell'})} (2\pi)^2 \delta_\mathrm{D}(\boldsymbol\xi - \boldsymbol\xi') C_\kappa^{(\gamma)}(\boldsymbol\xi,\boldsymbol\xi') (2\pi)^2 \delta_\mathrm{D}(\boldsymbol\xi - \boldsymbol\xi') C_\kappa^{(\gamma)}(\boldsymbol\xi',\boldsymbol\xi) + \tau_\gamma(\boldsymbol\ell,\boldsymbol\ell')~.
\end{equation}

It is easy to see that the above double integral does not vanish if and only if the two regions in multipole space $\Omega_{\boldsymbol\ell}$ and $\Omega_{\boldsymbol\ell'}$ coincide. Since we eventually considered circularly symmetric regions in multipole space and the modulus $\ell$ of the multipole vector actually takes integer values, it follows that

\begin{eqnarray}
\Psi\left[ P_\kappa^{(\gamma)}(\boldsymbol\ell), P_\kappa^{(\gamma)}(\boldsymbol\ell') \right] &=& \delta_\mathrm{K}^{\ell\ell'}\frac{2(2\pi)^2}{(4\pi f_\mathrm{sky})^2}\int_{\Omega_{\boldsymbol\ell}}\frac{\d^2 \boldsymbol\xi}{m^2(\Omega_{\boldsymbol\ell})} (2\pi)^2 \delta_\mathrm{D}(0) \left[C_\kappa^{(\gamma)}(\boldsymbol\xi,\boldsymbol\xi)\right]^2 + \tau_\gamma(\boldsymbol\ell,\boldsymbol\ell') = 
\nonumber\\
&=&\delta_\mathrm{K}^{\ell\ell'}\frac{2(2\pi)^2}{4\pi f_\mathrm{sky}}\int_{\Omega_{\boldsymbol\ell}}\frac{\d^2 \boldsymbol\xi}{m^2(\Omega_{\boldsymbol\ell})}\left[C_\kappa^{(\gamma)}(\boldsymbol\xi,\boldsymbol\xi)\right]^2 + \tau_\gamma(\boldsymbol\ell,\boldsymbol\ell')~.
\end{eqnarray}
By extracting the function $C_\kappa^{(\gamma)}(\boldsymbol\xi,\boldsymbol\xi)$ out of the integral, we obtain

\begin{equation}\label{eqn:cov_shear4}
\Psi\left[ P_\kappa^{(\gamma)}(\boldsymbol\ell), P_\kappa^{(\gamma)}(\boldsymbol\ell') \right] \simeq \delta_\mathrm{K}^{\ell\ell'}\frac{2\pi}{f_\mathrm{sky}m(\Omega_{\boldsymbol\ell})} \left[C_\kappa^{(\gamma)}(\boldsymbol\ell,\boldsymbol\ell)\right]^2 + \tau_\gamma(\boldsymbol\ell,\boldsymbol\ell') = \delta_\mathrm{K}^{\ell\ell'}\frac{2\pi}{f_\mathrm{sky} m(\Omega_{\boldsymbol\ell})} \left[P_\kappa(\boldsymbol\ell) + \frac{\sigma_\gamma^2}{\bar n}\right]^2 + \tau_\gamma(\boldsymbol\ell,\boldsymbol\ell')~.
\end{equation}
In many circumstances it is assumed that $\Omega_{\boldsymbol\ell}$ is just the shell in multipole space included between $\ell$ and $\ell + \Delta\ell$, where it is required that $\Delta\ell/\ell \ll 1$. If this is the case, then $m(\Omega_{\boldsymbol\ell}) = 2\pi\ell\Delta\ell + \pi\Delta\ell^2 = \pi\Delta\ell(2\ell+\Delta\ell)$. It follows that the covariance can be written as

\begin{equation}\label{eqn:covariance_shear}
\Psi\left[ P_\kappa^{(\gamma)}(\boldsymbol\ell), P_\kappa^{(\gamma)}(\boldsymbol\ell') \right] \simeq \delta_\mathrm{K}^{\ell\ell'}\frac{2}{\Delta\ell(2\ell+\Delta\ell)f_\mathrm{sky}} \left[P_\kappa(\boldsymbol\ell)+ \frac{\sigma_\gamma^2}{\bar n}\right]^2 + \tau_\gamma(\boldsymbol\ell,\boldsymbol\ell')~.
\end{equation}
This corresponds, modulo the non-Gaussian part, to the error expression first derived by \cite{KA92.1} and used, e.g., by \cite{HU99.2}.

A similar procedure and the same approximations can be used for the flexion-based estimate of the convergence power spectrum. The only relevant difference consists in replacing the function $C_\kappa^{(\gamma)}(\boldsymbol\ell,\boldsymbol\ell')$ with

\begin{equation}
C_\kappa^{(\varphi)}(\boldsymbol\ell,\boldsymbol\ell') \equiv P_\kappa(\boldsymbol\ell) + \frac{\sigma_\varphi^2}{\bar n \ell\ell'}~,
\end{equation}
so that, eventually,

\begin{equation}\label{eqn:covariance_flexion}
\Psi\left[ P_\kappa^{(\varphi)}(\boldsymbol\ell), P_\kappa^{(\varphi)}(\boldsymbol\ell') \right] \simeq \delta_\mathrm{K}^{\ell\ell'}\frac{2}{\Delta\ell(2\ell+\Delta\ell)f_\mathrm{sky}} \left[P_\kappa(\boldsymbol\ell)+ \frac{\sigma_\varphi^2}{\bar n\ell^2}\right]^2 + \tau_\varphi(\boldsymbol\ell,\boldsymbol\ell')~,
\end{equation}
where

\begin{equation}
\tau_\varphi(\boldsymbol\ell,\boldsymbol\ell') = \frac{1}{(4\pi f_\mathrm{sky})^2\bar n^4}\int_{\Omega_{\boldsymbol\ell}}\frac{\d^2 \boldsymbol\xi}{m(\Omega_{\boldsymbol\ell})}\int_{\Omega_{\boldsymbol\ell'}}\frac{\d^2 \boldsymbol\xi'}{m(\Omega_{\boldsymbol\ell'})}\left\langle \hat\kappa_\varphi(\boldsymbol\xi) \hat\kappa_\varphi^*(\boldsymbol\xi) \hat\kappa_\varphi(\boldsymbol\xi') \hat\kappa_\varphi^*(\boldsymbol\xi')  \right\rangle_\mathrm{c}
\end{equation}
is the contribution to the covariance coming from the non-linear clustering of matter.

It now remains to be calculated the covariance of the convergence power spectrum estimator based on the cross-correlation of shear and flexion. The details in this case are more cumbersome to work out, hence we report here only the most important steps and the final result. As usual, the covariance can be defined as

\begin{eqnarray}\label{eqn:3}
\Psi\left[ P_\kappa^{(x)}(\boldsymbol\ell), P_\kappa^{(x)}(\boldsymbol\ell') \right] &\equiv& \left\langle P_\kappa^{(x)}(\boldsymbol\ell)P_\kappa^{(x)}(\boldsymbol\ell')\right\rangle - P_\kappa(\boldsymbol\ell)P_\kappa(\boldsymbol\ell') = 
\nonumber\\
&=&\frac{1}{4}\frac{1}{(4\pi f_\mathrm{sky})^2\bar n^4}\int_{\Omega_{\boldsymbol\ell}}\frac{\d^2 \boldsymbol\xi}{m(\Omega_{\boldsymbol\ell})}\int_{\Omega_{\boldsymbol\ell'}}\frac{\d^2 \boldsymbol\xi'}{m(\Omega_{\boldsymbol\ell'})}\left\langle \left[ \hat\kappa_\gamma(\boldsymbol\xi)\hat\kappa_\varphi^*(\boldsymbol\xi) + \hat\kappa_\varphi(\boldsymbol\xi) \hat\kappa_\gamma^*(\boldsymbol\xi) \right] \left[ \hat\kappa_\gamma(\boldsymbol\xi')\hat\kappa_\varphi^*(\boldsymbol\xi') + \hat\kappa_\varphi(\boldsymbol\xi') \hat\kappa_\gamma^*(\boldsymbol\xi') \right]     \right\rangle -
\nonumber\\
&-& \frac{\sigma_x^2}{\bar n\ell'}P_\kappa(\boldsymbol\ell) - \frac{\sigma_x^2}{\bar n\ell}P_\kappa(\boldsymbol\ell') - \frac{\sigma_x^4}{\bar n^2\ell\ell'} - P_\kappa(\boldsymbol\ell)P_\kappa(\boldsymbol\ell')~.
\end{eqnarray}
The ensemble average inside the double integral above returns the sum of four ensemble averages, that thanks to the Wick's theorem provide a total of sixteen terms. The terms that do not mix functions depending on $\boldsymbol\xi$ with those depending on $\boldsymbol\xi'$ can be rearranged to cancel the last row of Eq. (\ref{eqn:3}). The remaining terms can be collected together by using the assumptions above, and return the following final expression:

\begin{equation}\label{eqn:covariance_cross}
\Psi \left[ P_\kappa^{(x)}(\boldsymbol\ell), P_\kappa^{(x)}(\boldsymbol\ell') \right] \simeq \delta_\mathrm{K}^{\ell\ell'}\frac{1}{\Delta\ell(2\ell+\Delta\ell)f_\mathrm{sky}} \left\lbrace \left[P_\kappa(\boldsymbol\ell)+ \frac{\sigma_\gamma^2}{\bar n}\right]\left[P_\kappa(\boldsymbol\ell)+ \frac{\sigma_\varphi^2}{\bar n\ell^2}\right] + \left[P_\kappa(\boldsymbol\ell)+ \frac{\sigma_x^2}{\bar n\ell}\right]^2\right\rbrace + \tau_x(\boldsymbol\ell,\boldsymbol\ell')~,
\end{equation}
where

\begin{eqnarray}
\tau_x(\boldsymbol\ell,\boldsymbol\ell')&\equiv& \frac{1}{4}\frac{1}{(4\pi f_\mathrm{sky})^2\bar n^4}\int_{\Omega_{\boldsymbol\ell}}\frac{\d^2 \boldsymbol\xi}{m(\Omega_{\boldsymbol\ell})}\int_{\Omega_{\boldsymbol\ell'}}\frac{\d^2 \boldsymbol\xi'}{m(\Omega_{\boldsymbol\ell'})} \left[\left\langle \hat\kappa_\gamma(\boldsymbol\xi) \hat\kappa_\varphi^*(\boldsymbol\xi) \hat\kappa_\gamma(\boldsymbol\xi') \hat\kappa_\varphi^*(\boldsymbol\xi')\right\rangle_\mathrm{c} + \left\langle \hat\kappa_\gamma(\boldsymbol\xi) \hat\kappa_\varphi^*(\boldsymbol\xi) \hat\kappa_\varphi(\boldsymbol\xi') \hat\kappa_\gamma^*(\boldsymbol\xi')\right\rangle_\mathrm{c} + \right.
\nonumber\\
&+&\left.\left\langle \hat\kappa_\varphi(\boldsymbol\xi) \hat\kappa_\gamma^*(\boldsymbol\xi) \hat\kappa_\gamma(\boldsymbol\xi') \hat\kappa_\varphi^*(\boldsymbol\xi')\right\rangle_\mathrm{c} +
\left\langle \hat\kappa_\varphi(\boldsymbol\xi) \hat\kappa_\gamma^*(\boldsymbol\xi) \hat\kappa_\varphi(\boldsymbol\xi') \hat\kappa_\gamma^*(\boldsymbol\xi')\right\rangle_\mathrm{c}  \right]
\end{eqnarray}
encapsulates the contribution due to the non-linear clustering of matter.

%%%%%%%%%%%%%%%%%%%%%%%%%%%%%%%%%%%%%%%%%%%%%%%%%%%
\subsection{Fisher matrices}
%%%%%%%%%%%%%%%%%%%%%%%%%%%%%%%%%%%%%%%%%%%%%%%%%%%

In order to compute the Fisher matrix for a generic estimator of the convergence power spectrum, let us assume that such estimator is measured for all multipole values included between $\ell_\mathrm{min}$ and $\ell_\mathrm{max}$, and let us leave $\Delta\ell$ unspecified. Under the assumption of a multivariate Gaussian likelihood the Fisher matrix can then be derived from the covariance as

\begin{equation}\label{eqn:fisher}
F_{jk} = \frac{1}{2}\mathrm{Tr}\left[ A_j(\boldsymbol\ell,\boldsymbol\ell')A_k(\boldsymbol\ell,\boldsymbol\ell') + \Psi^{-1}(\boldsymbol\ell,\boldsymbol\ell') M_{jk}(\boldsymbol\ell,\boldsymbol\ell')\right]~,
\end{equation}
where the moduli of both $\boldsymbol\ell$ and $\boldsymbol\ell'$ run from $\ell_\mathrm{min}$ to $\ell_\mathrm{max}$, and

\begin{equation}
A_j(\boldsymbol\ell,\boldsymbol\ell') = \Psi^{-1}(\boldsymbol\ell,\boldsymbol\ell')\frac{\partial}{\partial q_j} \Psi(\boldsymbol\ell,\boldsymbol\ell')~,
\end{equation}
while

\begin{equation}
M_{jk}(\boldsymbol\ell,\boldsymbol\ell') = \frac{\partial}{\partial q_j} P_\kappa(\boldsymbol\ell)\frac{\partial}{\partial q_k} P_\kappa(\boldsymbol\ell') +  \frac{\partial}{\partial q_k} P_\kappa(\boldsymbol\ell)\frac{\partial}{\partial q_j} P_\kappa(\boldsymbol\ell')~.
\end{equation}
The vector $\boldsymbol q$ represents the set of cosmological parameters on which the convergence power spectrum is assumed to depend on.

In order to be more specific, let us write down the Fisher matrices for the various convergence estimates considered in this Appendix. The Fisher matrix for the shear-based estimate easily follows from

\begin{equation}
A_j^{(\gamma)}(\boldsymbol\ell,\boldsymbol\ell') =  \frac{2\delta_\mathrm{K}^{\ell\ell'}}{P_\kappa(\boldsymbol\ell)+\sigma_\gamma^2/\bar n} \frac{\partial}{\partial q_j}P_\kappa(\boldsymbol\ell)~.
\end{equation}
It should be noted that the trace of the matrix in Eq. (\ref{eqn:fisher}) is just a sum over the diagonal terms, for which $\boldsymbol\ell = \boldsymbol\ell'$. It follows that the Fisher matrix can be simplified to

\begin{equation}\label{eqn:fisher_shear}
F^{(\gamma)}_{jk} = \sum_{\ell = \ell_\mathrm{min}}^{\ell_\mathrm{max}} \frac{4+\Delta\ell\left( 2\ell + \Delta\ell \right)f_\mathrm{sky}}{\left[P_\kappa(\boldsymbol\ell)+\sigma_\gamma^2/\bar n\right]^2} \frac{\partial}{\partial q_j} P_\kappa(\boldsymbol\ell)\frac{\partial}{\partial q_k} P_\kappa(\boldsymbol\ell)~.
\end{equation}
A similar expression holds for the Fisher matrix related to the flexion-based estimate of the convergence power spectrum, namely

\begin{equation}\label{eqn:fisher_flexion}
F^{(\varphi)}_{jk} = \sum_{\ell = \ell_\mathrm{min}}^{\ell_\mathrm{max}} \frac{4+\Delta\ell\left( 2\ell + \Delta\ell \right)f_\mathrm{sky}}{\left[P_\kappa(\boldsymbol\ell)+\sigma_\varphi^2/(\bar n \ell^2)\right]^2} \frac{\partial}{\partial q_j} P_\kappa(\boldsymbol\ell)\frac{\partial}{\partial q_k} P_\kappa(\boldsymbol\ell)~.
\end{equation}
An expression that, for high multipoles, is equivalent to this, has been derived along a different route by \cite{CA11.3}.

The Fisher matrix for the convergence power spectrum estimator based on the cross-correlation of shear and flexion is a little bit more cumbersome, but straightforward to compute nevertheless. The final result can be written as

\begin{eqnarray}\label{eqn:fisher_cross}
F^{(x)}_{jk} &=& \sum_{\ell = \ell_\mathrm{min}}^{\ell_\mathrm{max}} \frac{1}{\left[ P_\kappa(\boldsymbol\ell) + \sigma_\gamma^2/\bar n \right] \left[ P_\kappa(\boldsymbol\ell) + \sigma_\varphi^2/(\bar n\ell^2)\right] + \left[ P_\kappa(\boldsymbol\ell) + \sigma_x^2/(\bar n\ell) \right]^2} \left\lbrace 2\Delta\ell(2\ell+\Delta\ell)f_\mathrm{sky} + \phantom{\frac{\sigma_\gamma^2}{\sigma_\gamma^2}}\right.
\nonumber\\
&+&\left.\frac{16\left[ P_\kappa(\boldsymbol\ell) + \sigma_\gamma^2/(4\bar n) + \sigma_\varphi^2/(4\bar n\ell^2) + \sigma_x^2/(2\bar n\ell)\right]^2}{\left[ P_\kappa(\boldsymbol\ell) + \sigma_\gamma^2/\bar n \right] \left[ P_\kappa(\boldsymbol\ell) + \sigma_\varphi^2/(\bar n\ell^2)\right] + \left[ P_\kappa(\boldsymbol\ell) + \sigma_x^2/(\bar n\ell) \right]^2} \right\rbrace \frac{\partial}{\partial q_j} P_\kappa(\boldsymbol\ell)\frac{\partial}{\partial q_k} P_\kappa(\boldsymbol\ell)~.
\end{eqnarray}
These Fisher matrices can now be combined together to provide forecasted constraints on PNG.

\end{document}